\documentclass[a4paper,11pt]{article}
\pdfoutput=1 

\usepackage{jcappub} 

\usepackage[T1]{fontenc} 
\makeatletter
\def\@fpheader{\relax}
\makeatother

\title{\boldmath Palatini Higgs and Coleman-Weinberg inflation with non-minimal coupling}

\author[]{Nilay Bostan}

\affiliation{Department of Physics and Astronomy, University of Iowa, \\52242, Iowa City, IA, USA }

\emailAdd{nilay-bostan@uiowa.edu}

\abstract{We present the impact of non-minimal coupling $\xi \phi^2 R$ on the inflationary parameters by taking into account the models of single-field inflation with the inflaton that has a non-zero vacuum expectation value ($v$) after the period of inflation in Palatini gravity. We discuss the well-known symmetry breaking type potentials, namely the Higgs potential and Coleman-Weinberg potential. We show the inflationary predictions of these potentials, for both $\phi>v$ and $\phi<v$ inflation, the regions in the $v-\xi$ plane for which the values of $n_s$ and $r$ are in agreement with the recent measurements. We also show the linear inflation behavior as a solution of Coleman-Weinberg potential for $\xi v^2=1$ limit. Finally, we take into account the inflationary predictions of Coleman-Weinberg potential for preferred $\xi$ values as a function of $v$ in Palatini formalism. }

\begin{document}
\maketitle
\flushbottom

\section{Introduction}
\label{sec:intro}
Inflation \cite{Guth:1980zm,Linde:1981mu,Albrecht:1982wi,Linde:1983gd} is an accelerated expansion era thought to occur in the very early universe. It is considered that the hypothesis of cosmic inflation ensures a plausible clarification to some problems of the early universe such as the large scale homogeneity and isotropy of the universe, as well as, of the primordial density perturbations which evolve in cosmic structure. The idea that the inflationary scenario describes is based on slow-rolling scalar field, $\phi$, which is the so-called inflaton, over a flat potential $V(\phi)$. Many inflationary models have been debated in literature (see for an extensive subset \cite{Martin:2013tda}) and in general, they have been determined by the inflaton until now.

The inflationary parameters, in particular the spectral index $n_s$ and the tensor-to-scalar ratio $r$, of these models have been computed and compared with the constraints from the cosmic microwave background radiation (CMBR) \cite{Aghanim:2018eyx,Akrami:2018odb} temperature anisotropies and polarization measurements which have become more precise in recent years. The recent Keck Array/BICEP2 and Planck collaborations data especially constrain strongly the tensor-to-scalar ratio $r$ that provides reasonable explanation to the scale of inflation and the amplitude of primordial gravitational waves. From the latest data, $n_s=0.9649 \pm 0.0042$ as well as $r$ has an upper bound $r<0.056$ \cite{Akrami:2018odb}. The other parameter which is related with the inflation, namely the running of the spectral index, $\alpha=0.002 \pm 0.010$. For now, the recent constraints on the running of the spectral index are not satisfactory to test the inflationary models. However, it is considered to improve through 21 cm line observations around the level of $\alpha=\mathcal{O} (10^{-3})$ \cite{Kohri:2013mxa,Basse:2014qqa,Munoz:2016owz}. The inflationary parameters which are defined above are constrained at the pivot scale $k_* = 0.002$ Mpc$^{-1}$. 

In general, the calculations that are made under the assumption of the inflaton is minimally coupled. A well-known example is the scenario of inflation where the Standard Model Higgs boson acts as the inflaton with minimal coupling ($\xi=0$). However, in curved space-time, the non-minimal coupling $\xi \phi^2 R$ is necessary between the Ricci scalar and the inflaton for a renormalizable scalar field theory \cite{Callan:1970ze, Freedman:1974ze, Buchbinder:1992rb} and in spite of $\xi=0$ at the classical level, it is generated by quantum corrections \cite{Callan:1970ze}. Furthermore, depending upon $\xi$, inflationary parameters can change and it directly affects whether inflation occurs or not for the model under consideration \cite{Abbott:1981rg,Spokoiny:1984bd,Lucchin:1985ip,Futamase:1987ua,Fakir:1990eg,Salopek:1988qh,Amendola:1990nn,Faraoni:1996rf,Faraoni:2004pi}. Therefore, in this work, we take into account the inflation models in the existence of the non-minimal coupling to gravity. In literature, a large number of papers discuss the inflation with non-minimal coupling in Metric formulation \cite{Bezrukov:2010jz, Bostan:2018evz, Bezrukov:2007ep}. In this study, we discuss the non-minimally coupled Higgs potential and Coleman-Weinberg potential in Palatini formulation. For these potentials, we will present the effects of $\xi$ value on the inflationary parameters by taking into account that the inflaton has a non-zero vacuum expectation value $(v)$ after inflationary era. In the very early universe, some potentials can be related with symmetry breaking. This is the reason for taking into account a non-zero $v$ after inflation. Furthermore, non-minimally coupled inflationary models can not be clarified just with the potential form. The definition of degrees of freedom is necessary \cite{Bauer:2008zj}. 
In the metric formulation in general relativity (GR) \cite{Padmanabhan:2004fq,Paranjape:2006ca}, the metric and its first derivatives are the independent variables. On the contrary, in Palatini formulation  \cite{Attilio, Einstein, Ferraris}, the metric and the connection are the independent variables. For a given Einstein-Hilbert Lagrangian, these two formulations have the same equations of motion and, thus they represent equivalent physical theories. On the other hand, in the existence of non-minimal couplings between matter and gravity, such physical equivalence is lost and these two formulations correspond to two different theories of gravity (see for such refs. \cite{Bauer:2008zj,York:1972sj,Tenkanen:2017jih,Rasanen:2017ivk,Racioppi:2017spw,Tamanini:2010uq}). In particular for large $\xi$ values in Metric formulation, the attractor behavior, which is known as $\xi$-attractor models, occurs to the Starobinsky model. However, in Palatini formalism, this behavior is lost \cite{Kallosh:2013tua, Jarv:2017azx} as well as $r$ can take much smaller values according to the Metric formalism for large $\xi$ values \cite{Racioppi:2017spw,Barrie:2016rnv,Kannike:2015kda,Artymowski:2016dlz}. In literature, non-minimally coupled inflationary models in Palatini gravity are discussed in detail \cite{Tenkanen:2017jih,Bauer:2008zj,Rasanen:2017ivk,Jinno:2019und,Rubio:2019ypq,Enckell:2018kkc, Bostan:2019wsd, Tenkanen:2019jiq,Jarv:2020qqm}. Refs. \cite{Bauer:2008zj,Rasanen:2017ivk,Jinno:2019und,Rubio:2019ypq,Enckell:2018kkc, Bostan:2019wsd, Markkanen:2017tun} studied the Higgs potential in Palatini formulation. In ref. \cite{Bauer:2008zj}, self-interaction potential $V(\phi)$ is discussed for both Metric and Palatini formulations. In the large field limit, they found $n_s\simeq 0.968$ and $r\simeq10^{-14}$ in Palatini approach. In addition to this, Palatini Higgs inflation is analyzed in ref. \cite{Rasanen:2017ivk}. They obtained that $r$ is extremely suppressed, $1 \times 10^{-13}<r<2\times 10^{-5}$ in Palatini formulation. On the other hand, Coleman-Weinberg inflation in Palatini formulation is debated in refs. \cite{Racioppi:2017spw, Antoniadis:2018ywb, Gialamas:2020snr,Karam:2018mft}. According to ref. \cite{Racioppi:2017spw}, to generate the Planck scale, the vacuum expectation value $(v)$ of the inflaton must be $v^{2}=M^{2}_P/ \xi$ and they obtained when $\xi$ increases, $r$ decreases and saturating the linear limit for $\xi\gtrsim10^{-1}$ and they also showed for $\xi\gtrsim1$ values, Metric and Palatini formulations distinguish from each other. The predictions of Palatini formalism are smaller (larger) values of $r \ (n_s)$ compared to the Metric one as well as for $\xi\simeq1$, $r\simeq0.075$ in Palatini formulation \cite{Racioppi:2017spw}. Furthermore, ref. \cite{Gialamas:2020snr} analyzed the Coleman-Weinberg Palatini inflation by considering $v^{2}=M^{2}_P/ \xi$, same as ref. \cite{Racioppi:2017spw}. 

In our work, for non-minimally coupled Higgs and Coleman-Weinberg potentials in Palatini gravity, we show the inflationary parameters for preferred $\xi$ values in the regions $v-\xi$ for which $n_s$ and $r$ values that fit the current measurements. Furthermore, for Coleman-Weinberg potential, we present the inflationary parameters as functions of vacuum expectation values for chosen $\xi$ values. The predictions of observable parameters as functions of $v$ for different $\xi$ values for Higgs potential in Palatini formulation are analyzed in \cite{Bostan:2019wsd} with details. The paper is organized as follows: after a general explanation of inflation with non-minimal coupling in Palatini formulation (section \ref{general}), we discuss in detail two symmetry breaking potentials, namely the Higgs potential (section \ref{higgs}) and the Coleman-Weinberg potential (section \ref{cw}) in Palatini approach. Finally, in section \ref{conc}, we discuss our results in the paper.
\section{Non-minimally coupled inflation in Palatini gravity}\label{general}
We begin by considering an action of the form  

\begin{equation}\label{nonminimal_action}
S_J = \int \mathrm{d}^4x \sqrt{-g}\left(\frac{1}{2}F(\phi) g^{\mu\nu}R_{\mu\nu}(\Gamma) - \frac{1}{2} g^{\mu\nu}\partial_{\mu}\phi\partial_{\nu}\phi - V_J(\phi) \right),
\end{equation}
where $g$ is the determinant of the space-time metric $g_{\mu \nu}$ and the subscript $J$ denotes that the action is described in a Jordan frame. $\phi$ is the inflaton and $V_J(\phi)$ is its Jordan frame potential. $F(\phi)$ is a non-minimal coupling function and the action which is defined in eq. (\ref{nonminimal_action}) consists of a canonical kinetic term, a non-minimally coupled scalar field and a potential $V_J(\phi)$.  $R_{\mu\nu}$ is the Ricci tensor and it is defined as follows
\begin{equation}\label{Riccitensor}
R_{\mu\nu}=\partial_{\sigma}\Gamma_{\mu \nu}^{\sigma}-\partial_{\mu}\Gamma_{\sigma \nu}^{\sigma}+\Gamma_{\mu \nu}^{\rho}\Gamma_{\sigma \rho}^{\sigma}-\Gamma_{\sigma \nu}^{\rho}\Gamma^{\sigma}_{\mu \rho}.
\end{equation}
The connection is defined as a function of the metric tensor in the metric formulation. It is so called the Levi-Civita connection  ${\bar{\varGamma}={\bar{\varGamma}}(g^{\mu\nu})}$
\begin{equation} \label{vargammametric}
\bar{\varGamma}_{\mu\nu}^{\lambda}=\frac{1}{2}g^{\lambda \rho} (\partial_{\mu}g_{\nu \rho}+\partial_{\nu}g_{\rho \mu}-\partial_{\rho}g_{\mu\nu}).
\end{equation}
On the contrary, $g_{\mu \nu}$ and the connection $\varGamma$ are treated as independent variables in the Palatini formulation, and it is assumed that the connection is torsion-free, i.e. $\varGamma_{\mu\nu}^{\lambda}=\varGamma_{\nu\mu }^{\lambda}$. By solving the EoM, it can be obtained in the form \cite{Bauer:2008zj}
\begin{equation}\label{vargammapalatini}
\Gamma^{\lambda}_{\mu\nu} = \overline{\Gamma}^{\lambda}_{\mu\nu}
+ \delta^{\lambda}_{\mu} \partial_{\nu} \omega(\phi) +
\delta^{\lambda}_{\nu} \partial_{\mu} \omega(\phi)\nonumber\\ - g_{\mu \nu} \partial^{\lambda} \omega(\phi),
\end{equation}
where 
\begin{eqnarray}
\label{omega}
\omega\left(\phi\right)=\ln\sqrt{F(\phi)},
\end{eqnarray}
in the Palatini formalism. In this paper, we will calculate the inflationary parameters of symmetry-breaking type potentials. We consider that the $F(\phi)$ consists of a constant term: $m^2$ and a non-minimal coupling term: $\xi \phi^2 R$. We will use units, where the reduced Planck scale $m_P=1/\sqrt{8\pi
G}\approx2.4\times10^{18}\text{ GeV}$ is set equal to unity. Therefore, either $F(\phi)\to1$ or $\phi\to0$ is necessary after inflation. As a result, by taking $m^2=1-\xi v^2$, we find $F(\phi)=m^2+\xi \phi^2=1+\xi(\phi^2-v^2) $ \cite{Bostan:2018evz,Bostan:2019wsd}. By using $F(\phi)=1+\xi(\phi^2-v^2) $, we will analyze the inflationary parameters for the Higgs potential and the Coleman-Weinberg potential, of the inflaton field values for $\phi>v$ and $\phi<v$.

\subsection{Overview of inflationary parameters}
To calculate the inflationary parameters for eq. (\ref{nonminimal_action}), it is suitable to switch to the Einstein (E) frame by using a Weyl rescaling $g_{E, \mu \nu}=g_{\mu \nu}/F(\phi)$, thus the Einstein frame action can be found in the form
\begin{eqnarray}\label{einsteinframe}
S_E = \int \mathrm{d}^4x \sqrt{-g_{E}}\left(\frac{1}{2}g_E^{\mu\nu}R_{E, \mu \nu}(\Gamma)-\frac{1}{2Z(\phi)}\, g_E^{\mu\nu} \partial_{\mu}\phi\partial_{\nu}\phi - \frac{V_E(\phi)}{F(\phi)^2} \right),
\end{eqnarray}
where
\begin{equation} \label{Zphi}
Z^{-1}(\phi)=\frac{1}{F(\phi)},
\end{equation}
in the Palatini formalism. If a field redefinition is made by using
\begin{equation}\label{redefine}
\mathrm{d}\chi=\frac{\mathrm{d}\phi}{\sqrt{Z(\phi)}}\,,
\end{equation}
we find the action with a canonical kinetic term for a minimally coupled scalar field $\chi$.  By taking into account the eq. \eqref{redefine}, the action in the Einstein frame in terms of $\chi$ can be achieved as follows
\begin{eqnarray}\label{einsteinframe}
S_E = \int \mathrm{d}^4x \sqrt{-g_{E}}\left(\frac{1}{2}g_E^{\mu\nu}R_{E}(\Gamma)-\frac{1}{2}\, g_E^{\mu\nu} \partial_{\mu}\chi\partial_{\nu}\chi - V_E(\chi) \right).
\end{eqnarray}
As a result, if the Einstein frame potential can be written in terms of the canonical scalar field $\chi$, inflationary parameters can be obtained by using the slow-roll parameters
\cite{Lyth:2009zz}
\begin{equation}\label{slowroll1}
\epsilon =\frac{1}{2}\left( \frac{V_{\chi} }{V}\right) ^{2}\,, \quad
\eta = \frac{V_{\chi\chi} }{V}  \,, \quad
\xi ^{2} = \frac{V_{\chi} V_{\chi \chi\chi} }{V^{2}}\,,
\end{equation}
here $\chi$'s in the subscripts denote derivatives. Inflationary parameters can be described with the slow-roll approximation in the form
\begin{eqnarray}\label{nsralpha1}
n_s = 1 - 6 \epsilon + 2 \eta \,,\quad
r = 16 \epsilon, \quad
\alpha=\frac{\mathrm{d}n_s}{\mathrm{d}\ln k} = 16 \epsilon \eta - 24 \epsilon^2 - 2 \xi^2\,.
\end{eqnarray}
In the slow-roll approximation, the number of e-folds is acquired as follows
\begin{equation} \label{efold1}
N_*=\int^{\chi_*}_{\chi_e}\frac{V\rm{d}\chi}{V_{\chi}}\,. \end{equation}
Here, the subscript ``$_*$'' represents the quantities which the scale
corresponding to $k_*$ exited the horizon, as well as $\chi_e$ is the inflaton
value at the end of inflation, which we calculate by $\epsilon(\chi_e) =
1$. 

The amplitude of the curvature perturbation in terms of $\chi$ is given in the form
\begin{equation} \label{perturb1}
\Delta_\mathcal{R}=\frac{1}{2\sqrt{3}\pi}\frac{V^{3/2}}{|V_{\chi}|},
\end{equation}
which should be compatible with $\Delta_\mathcal{R}^2\approx   2.1\times10^{-9}$ from the Planck results \cite{Aghanim:2018eyx} for the pivot scale $k_* = 0.002$ Mpc$^{-1}$.  

On the other hand, we need to redefine the slow-roll parameters in terms of the original field $\phi$ for our numerical calculations, since it is not simple to calculate the inflationary potential in terms of $\chi$ for general $\xi$ and $v$ values. By using eq. \eqref{redefine}, eq. \eqref{slowroll1} can be obtained in terms of $\phi$ \cite{Linde:2011nh}
\begin{equation}\label{slowroll2}  
\epsilon=Z\epsilon_{\phi}\,,\quad
\eta=Z\eta_{\phi}+{\rm sgn}(V')Z'\sqrt{\frac{\epsilon_{\phi}}{2}}\,,\quad
\xi^2=Z\left(Z\xi^2_{\phi}+3{\rm sgn}(V')Z'\eta_{\phi}\sqrt{\frac{\epsilon_{\phi}}{2}}+Z''\epsilon_{\phi}\right)\,,
\end{equation}
where we described 
\begin{equation}
\epsilon_{\phi} =\frac{1}{2}\left( \frac{V^{\prime} }{V}\right) ^{2}\,, \quad
\eta_{\phi} = \frac{V^{\prime \prime} }{V}  \,, \quad
\xi ^{2} _{\phi}= \frac{V^{\prime} V^{\prime \prime\prime} }{V^{2}}\,.
\end{equation}
Furthermore, eqs. \eqref{efold1} and \eqref{perturb1} can be found in terms of $\phi$ by using
\begin{eqnarray}\label{perturb2}
N_*&=&\rm{sgn}(V')\int^{\phi_*}_{\phi_e}\frac{\mathrm{d}\phi}{Z(\phi)\sqrt{2\epsilon_{\phi}}}\,,\\
\label{efold2} \Delta_\mathcal{R}&=&\frac{1}{2\sqrt{3}\pi}\frac{V^{3/2}}{\sqrt{Z}|V^{\prime}|}\,.
\end{eqnarray}

In order to calculate the inflationary parameters numerically, we need to obtain the $N_*$ value numerically as well. On the assumption that a standard thermal history after inflation, $N_*$ is given in the form \cite{Liddle:2003as}
\begin{equation} \label{efolds}
N_*\approx64.7+\frac12\ln\rho_*-\frac{1}{3(1+\omega_r)}\ln\rho_e
+\left(\frac{1}{3(1+\omega_r)}-\frac14\right)\ln\rho_r\,,
\end{equation}
here $\rho_{e}=(3/2)V(\phi_{e})$, at the end of inflation, it indicates the
energy density, $\rho_*\approx V(\phi_*)$ defines the energy density when the scales corresponding to $k_*$ exited the horizon, $\rho_r$, at the
end of reheating, is the energy density as well as during reheating, $\omega_r$ is the parameter of equation of state. We take $\omega_r$ to be a constant. The predictions of the inflationary parameters can change depending upon $N_*$. We can define three different cases for $N_*$ depending on the reheating temperature ($T_r$): in the high-$N$ scenario $\omega_r=1/3$, which corresponds to the assumption of instant reheating, in the middle-$N$ scenario $\omega_r=0$ and $T_r=10^{9}$ GeV, computing $\rho_r$ by using the Standard Model value for the number of relativistic degrees of freedom $g_*=106.75$. In the low-N scenario $\omega_{r}=0$, which is similar to the middle-$N$ scenario but the reheat temperature $T_r=100$ GeV. In this work, we will calculate the inflationary parameters in the high-$N$ case for both potentials which are considered.


\section{Higgs potential}\label{higgs}
In this section, we analyze the inflationary predictions of the Higgs potential which is associated with the symmetry-breaking \cite{Goldstone:1961eq}
\begin{equation}\label{higgs1}
V_J(\phi)=A \left[1-\left(\frac{\phi}{v}\right)^2\right]^2\,.
\end{equation}
The minimally coupled Higgs potential has been discussed comprehensively in literature \cite{Vilenkin:1994pv, Linde:1994wt,Destri:2007pv,Martin:2013tda,Okada:2014lxa}. Once inflation occurs near the minimum, the minimally coupled Higgs potential approximates the quadratic potential. Thus, the inflationary predictions of Higgs potential with minimal coupling in terms of $N_*$ are written in the form
\begin{equation}\label{quadratic}
n_s\approx1-\frac{2}{N_*}\,,\quad r\approx\frac{8}{N_*}\,,\quad
\alpha\approx-\frac{2}{N_*^2}.
\end{equation}
These results are attained for the cases of both $\phi>v$ and $\phi<v$. In this study, we consider the Higgs potential with non-minimal coupling in Palatini formulation. For this potential, we calculate the inflationary parameters in the $v-\xi$ plane inside the 95\% (68\%) confidence level contours given by the Keck Array/BICEP2 and Planck collaborations \cite{Ade:2018gkx} for the cases of $\phi>v$ and $\phi<v$.

The predictions of the Higgs potential with non-minimal coupling in Palatini approach is different from the minimal case. By using eq. \eqref{perturb2}, the number of e-folds can be obtained in the form
\begin{equation}\label{quad}
N_*=\frac{1}{8}\left( \phi_*^2-\phi_e^2\right)-\frac{v^2}{4}\ln\frac{\phi_*}{\phi_e}.
\end{equation}
For the large-field limit (during inflation: $\phi^2\gg v^2$), the inflationary parameters of the Higgs potential in Palatini formalism can be obtained as follows \cite{Bostan:2019wsd}
\begin{equation}\label{higgsns}
n_s\approx1-\frac{2}{N_*}, \qquad r\approx \frac{2}{\xi N^2_*}, \qquad \alpha\approx-\frac{2}{N_*^2}.
\end{equation}
However, when cosmological scales exit the horizon for the case of $\phi\ll v$, the potential can be written in the form effectively
\begin{equation}\label{higgsns1}
V_E(\phi)\approx A \left[1-2\left(\frac{\phi}{v}\right)^2\right].
\end{equation}
This form of potential approximates the small-field inflation type potentials  that also occur in some supersymmetric inflationary models \cite{Izawa:1996dv,Kawasaki:2003zv,Senoguz:2004ky}, for the case of $\phi<v$ during inflation.
The predictions of this potential, which is described in eq. \eqref{higgsns1}, are such that $r$ takes very tiny values and $n_s\approx1-8/v^2$. 
\begin{figure}[tbp]
\centering 
\includegraphics[width=.6\textwidth]{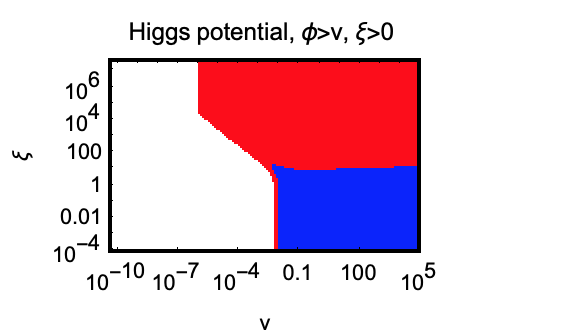}

\

\includegraphics[width=.496\textwidth]{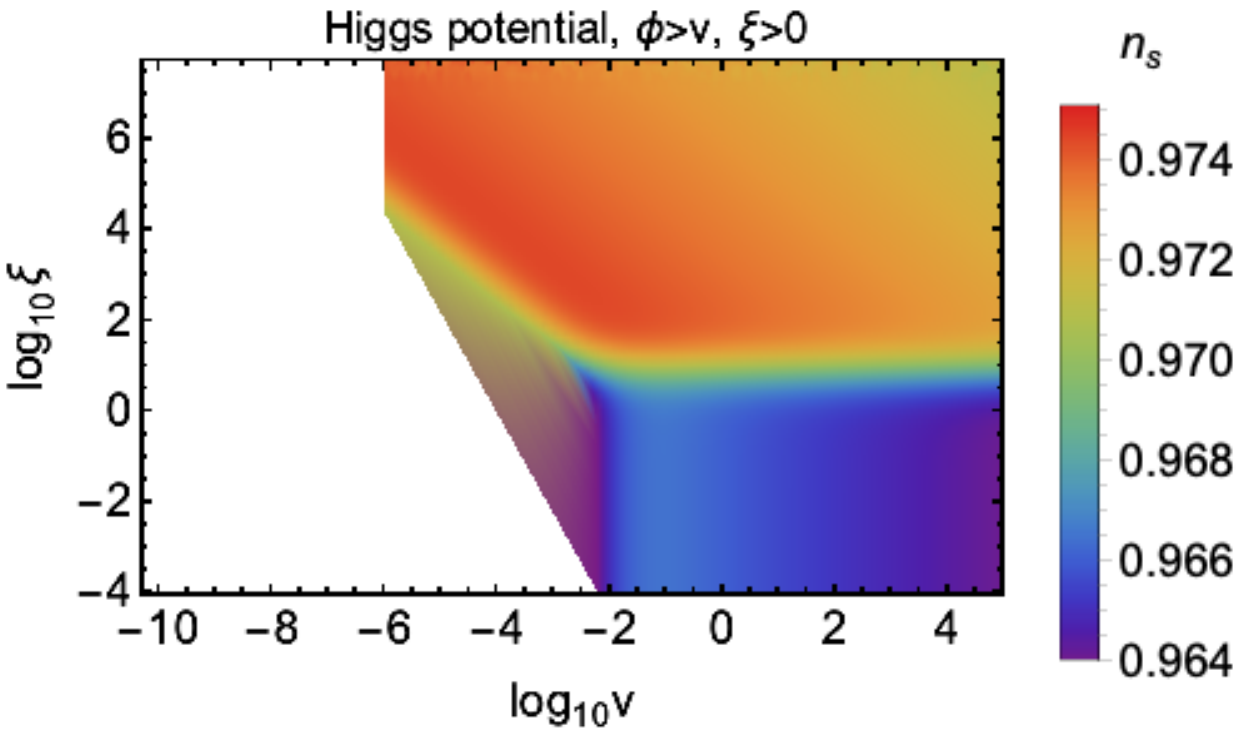}
\includegraphics[width=.496\textwidth]{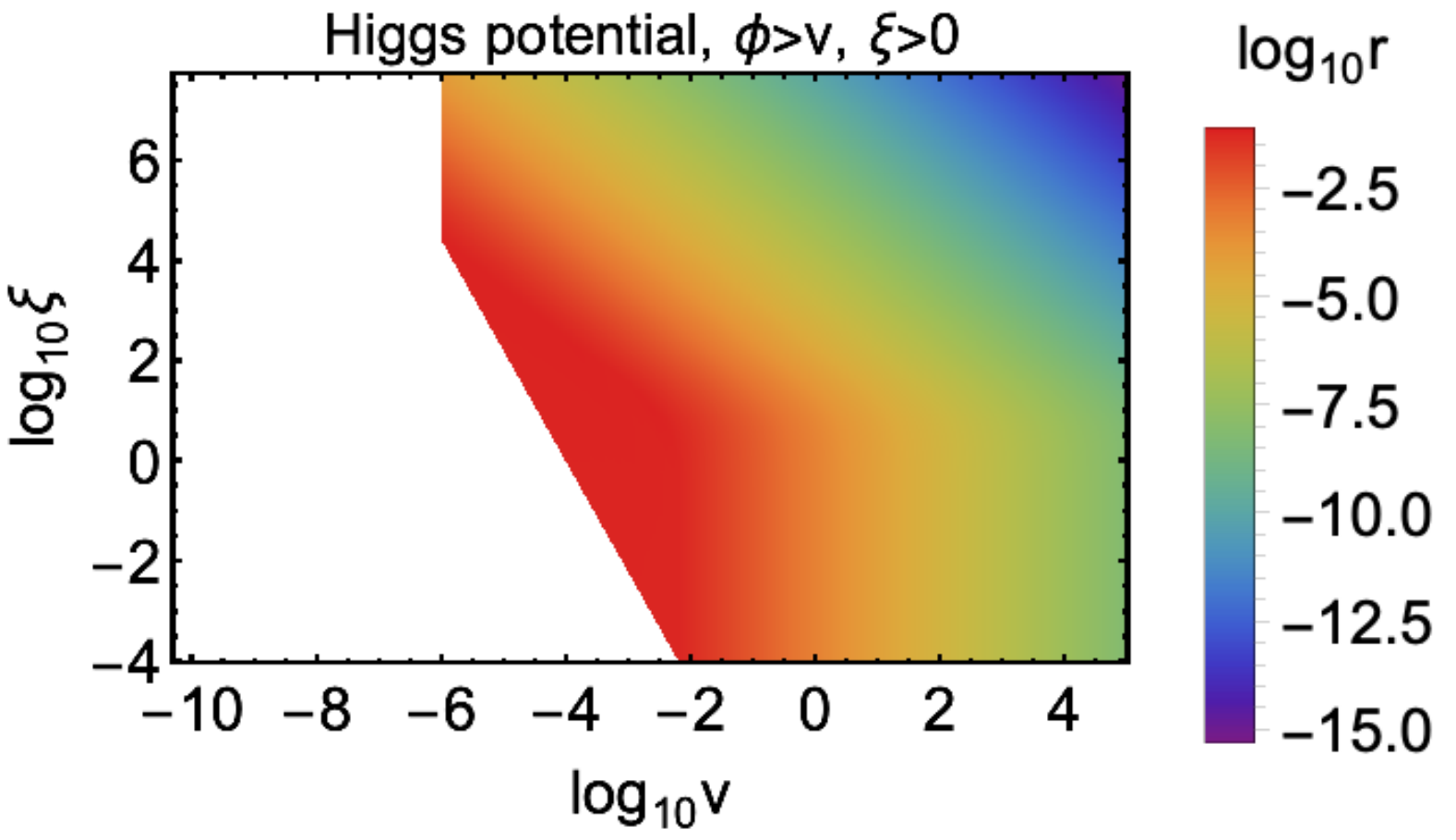}
\caption{\label{higgsabove} The results of the Higgs potential in Palatini formulation for the cases of $\phi>v$ and $\xi>0$. In the top figure, red (blue) display the regions in the $v-\xi$ plane that predict $n_s$ and $r$ values inside the 95\% (68\%) CL Keck Array/BICEP2 and Planck  contours \cite{Ade:2018gkx}. Bottom figures correspond to the
		$n_s$ and $r$ values in these regions.}
\end{figure}

\begin{figure}[tbp]
\centering 
\includegraphics[width=.6\textwidth]{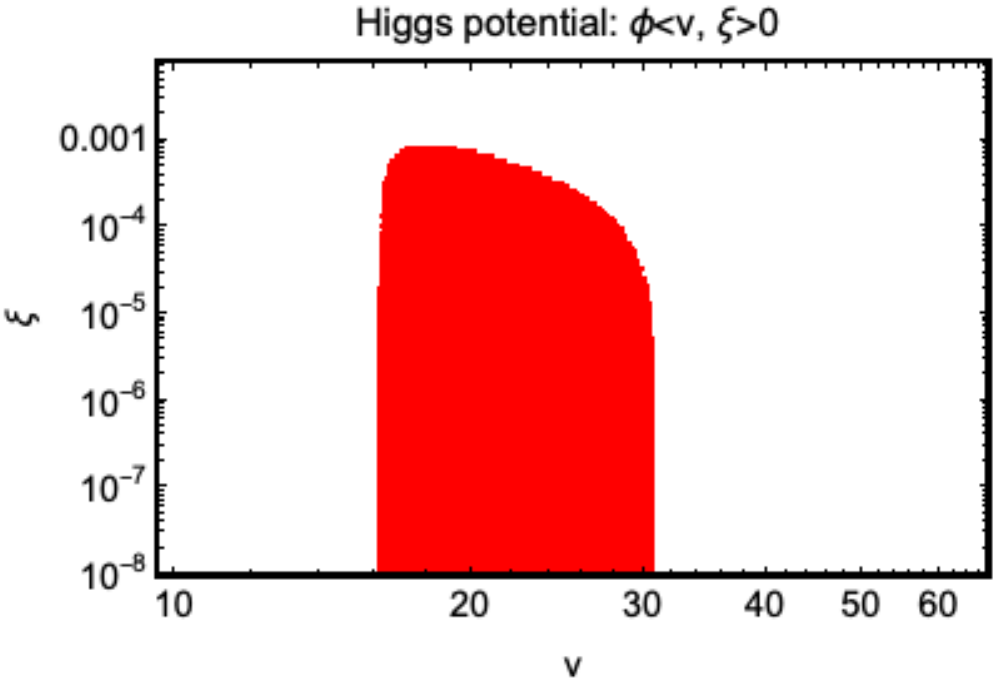}

\

\includegraphics[width=.496\textwidth]{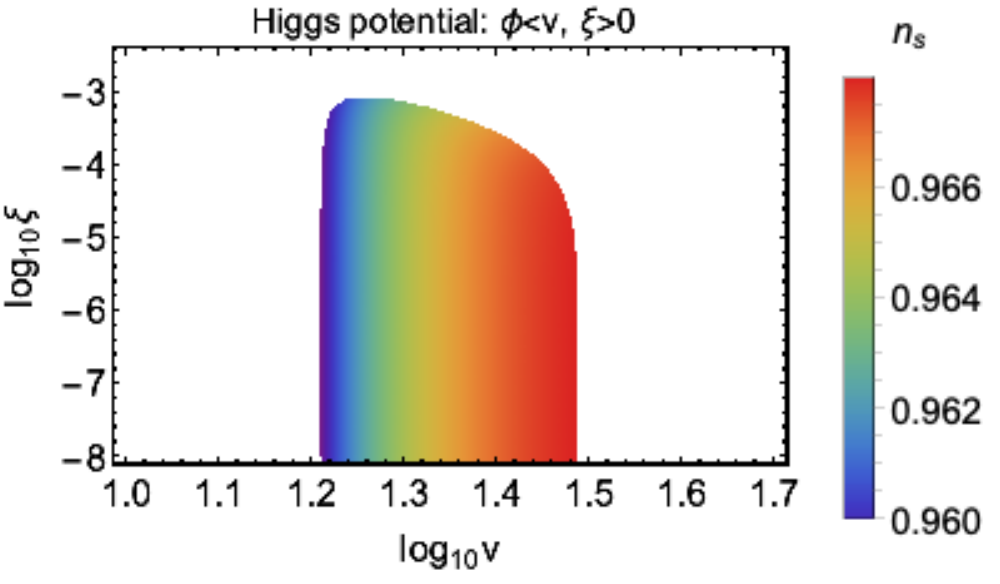}
\includegraphics[width=.496\textwidth]{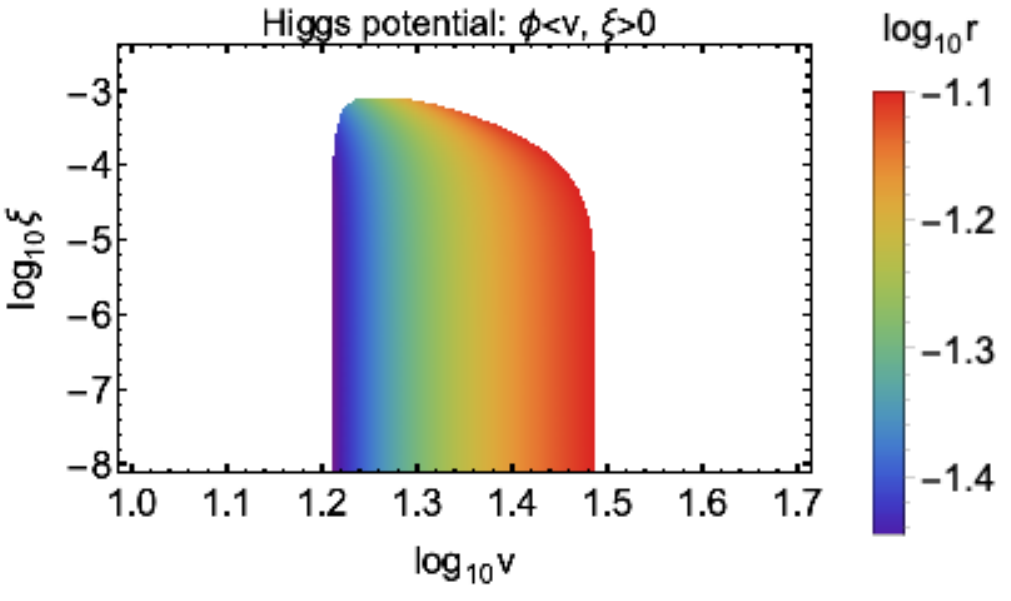}
\caption{\label{higgsbelowpos} The results of the Higgs potential in Palatini formulation for the cases of $\phi<v$ and $\xi>0$. In the top figure, the red displays the regions in the $v-\xi$ plane that predict $n_s$ and $r$ values inside the 95\% CL Keck Array/BICEP2 and Planck  contour \cite{Ade:2018gkx}. Bottom figures correspond to the
		$n_s$ and $r$ values in this region.}
\end{figure}
\begin{figure}[tbp]
\centering 
\includegraphics[width=.53\textwidth]{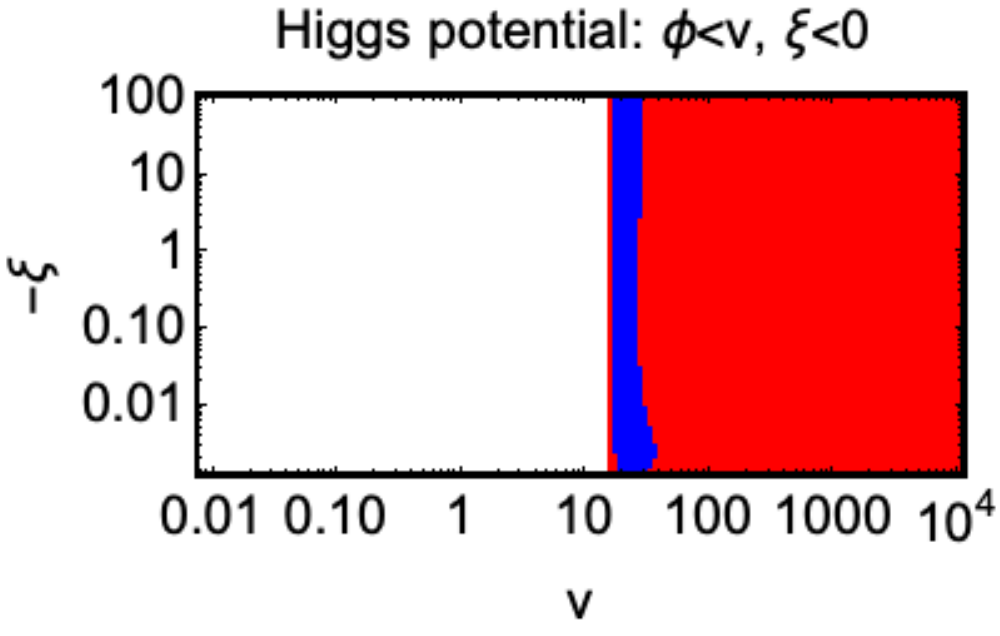}

\

\includegraphics[width=.496\textwidth]{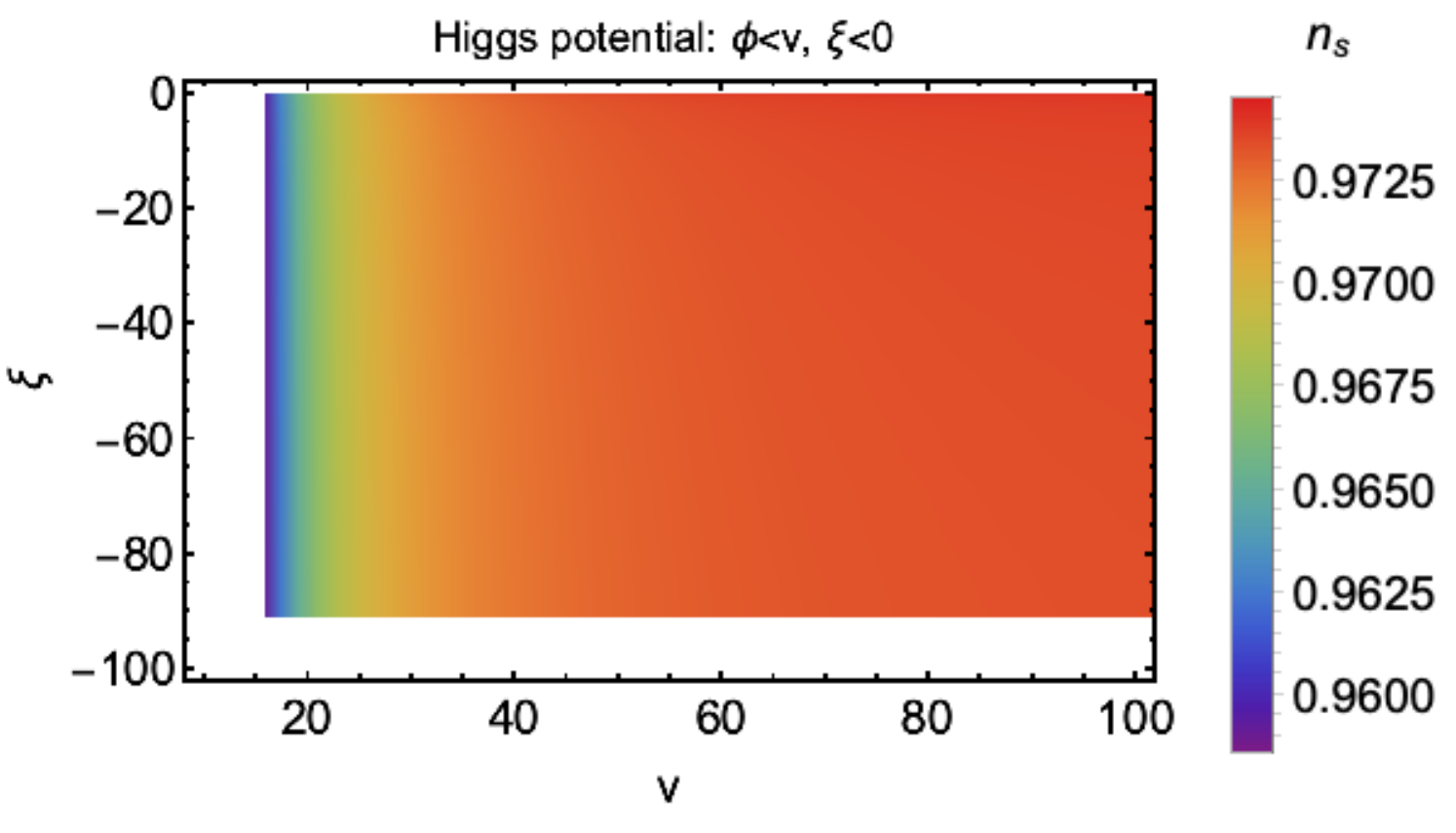}
\includegraphics[width=.496\textwidth]{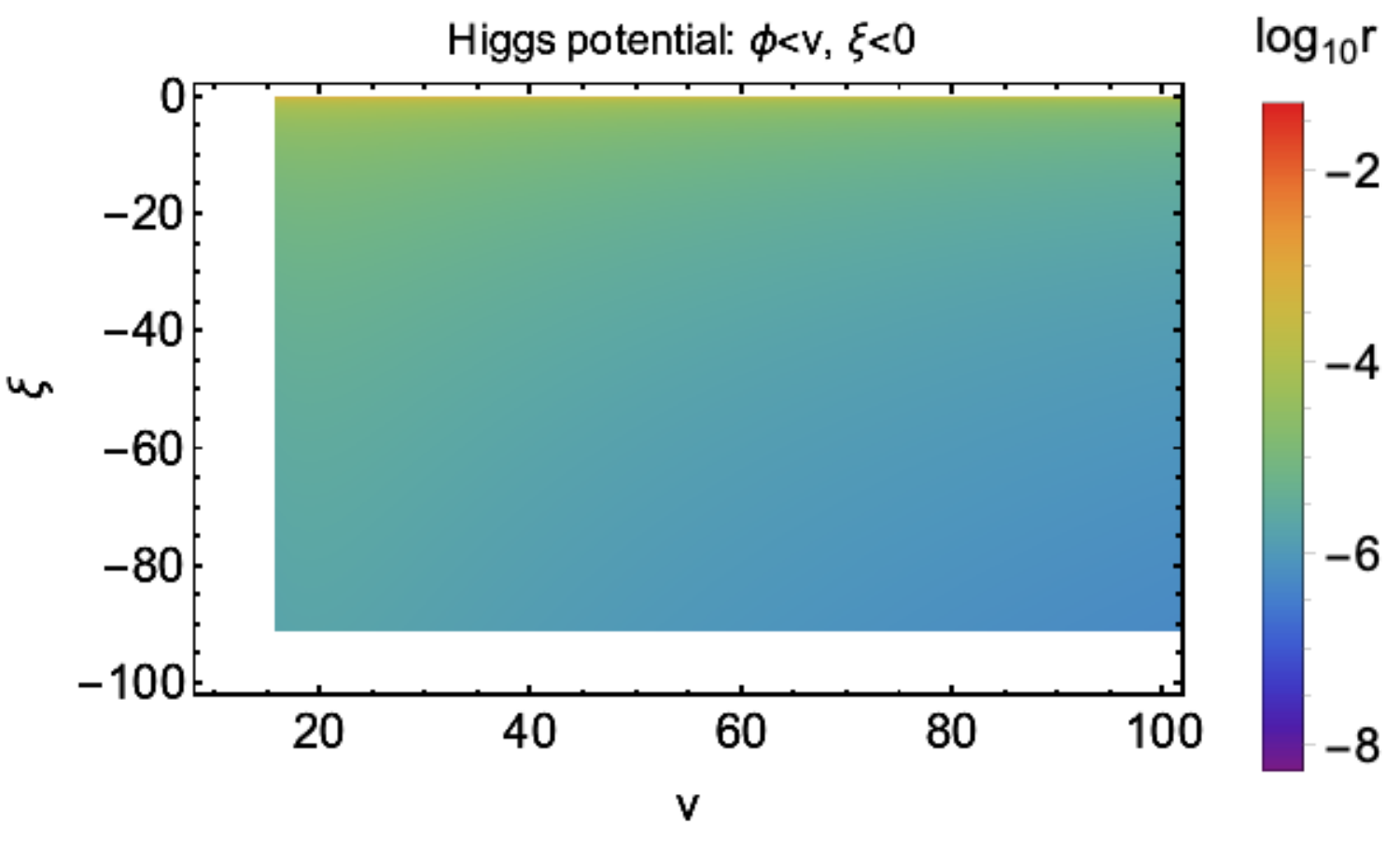}
\caption{The results of the Higgs potential in Palatini formulation for the cases of $\phi<v$ and $\xi<0$. In the top figure, the red (blue) display the regions in the $v-\xi$ plane that predict $n_s$ and $r$ values inside 95\% (68\%) CL Keck Array/BICEP2 and Planck  contours \cite{Ade:2018gkx}. Bottom figures correspond to the
		$n_s$ and $r$ values in these regions.\label{higgsbelowneg} }
\end{figure}
In the literature, \cite{Bauer:2008zj} compared the inflationary parameters of the Higgs potential for Palatini and Metric formalisms in the large-field limit ($\phi^2\gg v^2$) during inflation. The differences between Palatini and Metric formalisms are also analyzed for the Higgs potential in ref. \cite{Rasanen:2017ivk}. They used loop corrections with a simple approximation and they considered different parts of the potential, such as on plateau, at the point of the hilltop and the critical point, as well as in false vacuum. They showed that $r$ varies vastly for different parts of the potential, especially in the plateau, $r$ is very tiny and $1\times 10^{-13}<r<10^{-5}$. Furthermore, \cite{Jinno:2019und} studied the Higgs(-like) inflation with higher dimensional Weyl operators in the case of non-minimal couplings and with the potential for Palatini and Metric formulations. They presented that the inflationary parameters are stable towards the Weyl operators nearby the attractor point in Metric formalism, however, they are very sensitive in the Palatini approach. They analyzed the inflationary predictions by taking $F(\phi)=1+\xi \phi^2$ for $N_*=50$ and $N_*=60$ values. Ref. \cite{Rubio:2019ypq} analyzed the $n_s$, its amplitude and $r$ for the Higgs potential in Palatini gravity for $\xi \phi^2\gg1$ limit and they also discussed the details of the preheating stage in Palatini Higgs inflation by using $N_*\simeq 54.9-(1/4)\log\xi$. In addition to this, ref. \cite{Bostan:2019wsd,Markkanen:2017tun} also investigated the Palatini Higgs inflation. Ref. \cite{Markkanen:2017tun} investigated for the large-field limit, $N_*=60$ required $\xi=\mathcal{O}(10^4)$ in the Metric case and $\xi=\mathcal{O}(10^9)$ in the Palatini approach. They also showed for $N_*=60$, $r\simeq3\times 10^{-3}$ (Metric), but $r$ is highly suppressed to be inside the upcoming experiments in Palatini formalism. Ref. \cite{Bostan:2019wsd} calculated the inflationary parameters for Palatini Higgs inflation for $\phi>v$ and $\phi<v$ inflation just for selected a few $\xi$ values. Even though the Palatini Higgs inflation were previously discussed with details, there still are some shortcomings in the literature which we mention in this section. Unlike the literature, in this section, we present the inflationary predictions for non-minimally coupled Higgs potential in Palatini gravity for both $\phi>v$ and $\phi<v$ cases by taking account a non-zero $v$ after the inflationary era is that the potential can be related with symmetry breaking in the very early universe. We also compare our numerical results in the wide range of $v-\xi$ plane in the 95\% (68\%) CL contour with the data given by the Keck Array/BICEP2 and Planck collaborations \cite{Ade:2018gkx} for the assumption of a standard thermal history after inflation in the high-N case which we explained in the previous section. First of all, we present the results of $\phi>v$ and $\xi>0$ cases. It can be seen from fig. \ref{higgsabove} that, in the range of $10 \lesssim \xi \lesssim 10^7$ and $10^{-6} \lesssim v \lesssim 10^{5}$, the predictions are in the 95\% CL. On the other hand, for the range of $10^{-4} \lesssim \xi \lesssim 10$ and $10^{-2} \lesssim v \lesssim 10^{5}$, the predictions are in the 68\% CL. Furthermore, for the $\xi\gg1$ values, $r$ is very suppressed and $r \simeq 10^{-15}$. This result is presented in \cite{Bauer:2008zj, Bostan:2019wsd}. Ref. \cite{Bauer:2008zj} analyzed the inflationary predictions for Palatini Higgs potential in the large-field limit ($\xi \phi^2\gg1$) and they found $n_s\simeq0.97$ and $r\simeq 10^{-14}$. Our results for $\xi \phi^2\gg1$ limit, which we presented in fig. \ref{higgsabove}, are compatible with \cite{Bauer:2008zj}. As it can be seen from the bottom plots in fig. \ref{higgsabove}, we obtained for the values of $\xi\simeq10^6$ and $v\simeq10^{4}$, $n_s\simeq0.974$ and $r\simeq10^{-15}$. In addition to this, $\phi>v$ and $\xi<0$ cases are outside the observational range, since these cases lead to a larger $r$. Fig. \ref{higgsbelowpos} is plotted for the cases of $\phi<v$ and $\xi>0$. According to fig. \ref{higgsbelowpos}, the predictions of $n_s$ and $r$ for the range of $10^{-5}\lesssim \xi \lesssim 10^{-3}$ and $17\lesssim v \lesssim 32$, are in the 95\% CL contour. However, $\xi\gtrsim 10^{-3}$ and for all $v$ values, the predictions are ruled out. As it can be seen from the bottom plots in fig. \ref{higgsbelowpos}, the range is between $10^{-5}\lesssim \xi \lesssim 10^{-3}$ and $17\lesssim v \lesssim 32$, $0.96 \lesssim n_s \lesssim 0.967$ and $0.032\lesssim r \lesssim 0.079 $. Finally, we also analyzed the cases of $\phi<v$ and $\xi<0$, and the results can be seen in fig. \ref{higgsbelowneg}. The predictions of $n_s$ and $r$ can be inside both 68\% and 95\% CL. For the values of $-100\lesssim \xi \lesssim -10^{-3}$ and $10\lesssim v \lesssim 30$, the predictions are in 68\% CL but for the same $\xi$ ranges and $30\lesssim v \lesssim 10^{4}$, $n_s$ and $r$ are in 95\% CL.

\section{Coleman-Weinberg potential}\label{cw}
In this section, we take into account the mechanism of Coleman-Weinberg symmetry breaking type inflation \cite{Coleman:1973jx}, which was proposed in the early eighties. The effective potential is given as follows \cite{Albrecht:1984qt,Linde:2005ht}

\begin{equation}\label{cwpot}
V_J(\phi)=A \phi^4\left[\ln \left(\frac{\phi}{v}\right)-\frac{1}{4}\right]+\frac{A v^4}{4}.
\end{equation}
The Coleman-Weinberg potential with minimal coupling is analyzed in refs. \cite{Shafi:2006cs,Barenboim:2013wra,Kannike:2014mia,Senoguz:2015lba} in detail. The inflationary parameters of this potential are very similar to the results of the Higgs potential. For the case of $\phi>v$ inflation, the predictions of $n_s$ and $r$ interpolate between the quartic and quadratic potential limits and they are ruled out. On the other hand for $\phi<v$ inflation, the case of $v^2\ll 4N_*$ when cosmological scales exit the horizon, the potential approximates the small-field inflation potentials. 
\begin{figure}[tbp]
\centering 
\includegraphics[width=.8\textwidth]{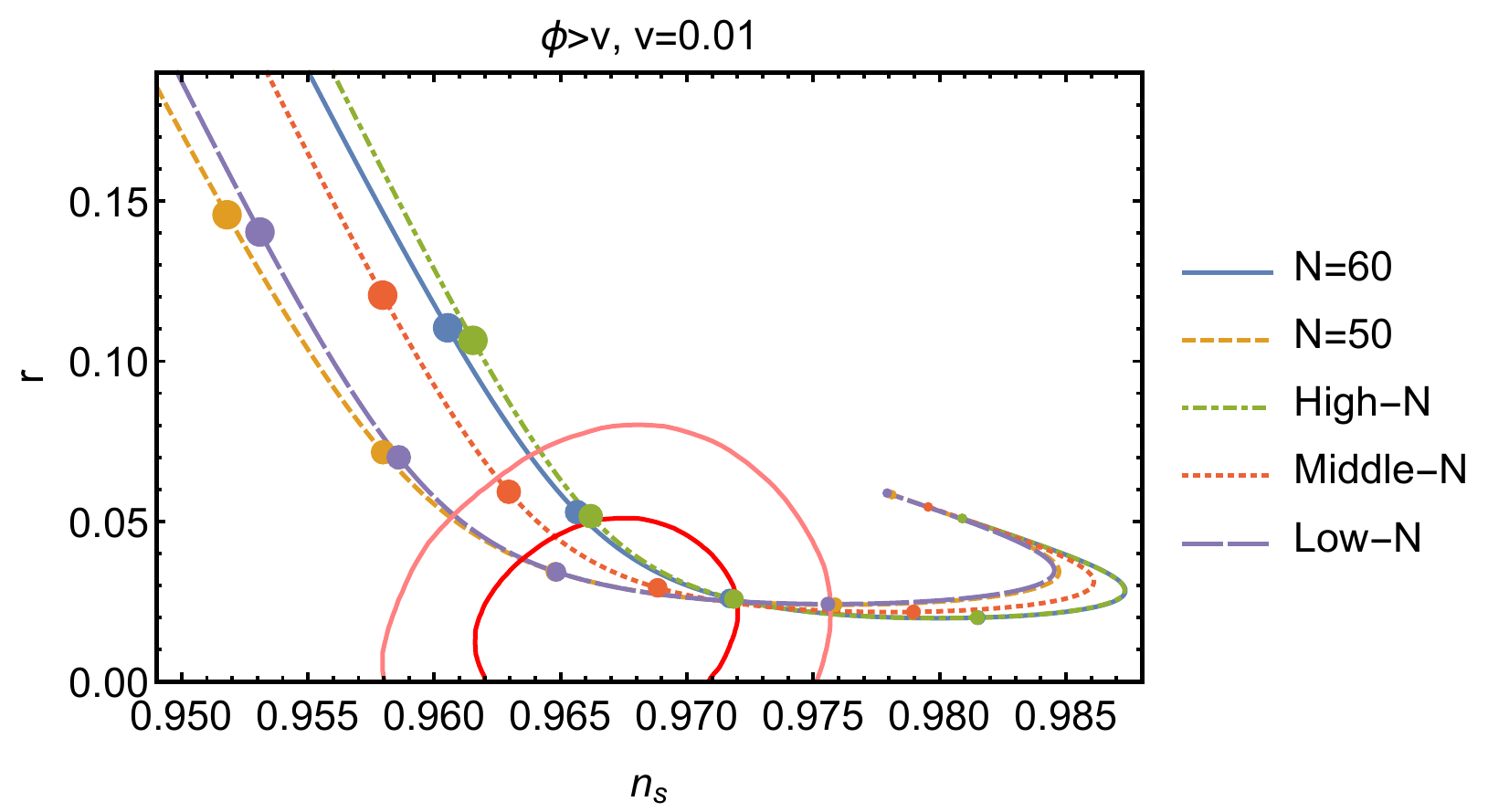}

\caption{The figure shows the $n_s$-$r$ predictions for varied $\xi$ values and $v=0.01$ for different reheating cases as clarified in the text for Coleman-Weinberg potential in Palatini formulation. The dots on the curves indicate $\xi=10^{-2.5},\,10^{-2},\,10^{-1.5},\,10^{-1},$ and $\,1$, from top to bottom. The pink
		(red) indicates the 95\% (68\%) CL Keck Array/BICEP2 and Planck contours  \cite{Ade:2018gkx}.}
	\label{fig1}
\end{figure}

The curves of $n_s-r$ for different $N_*$ cases, which are described in section \ref{general}, are displayed in figure \ref{fig1} for the Coleman-Weinberg potential in Palatini formalism with the CL Keck Array/BICEP2 and Planck contours  \cite{Ade:2018gkx}. 
\begin{figure}[tbp]
\centering 
\includegraphics[width=1.04\textwidth]{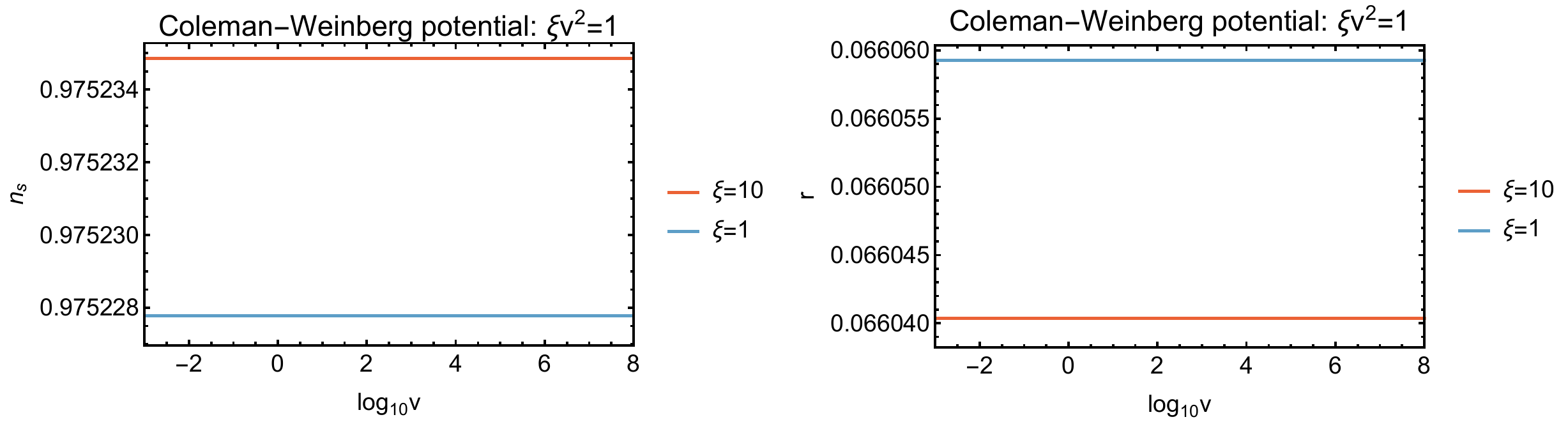}

\caption{\label{cwind} The $n_s-r$ predictions of Coleman-Weinberg potential in Palatini formulation in the case of $\xi v^2=1$ for $\xi=1$ and $\xi=10$ values.}
\end{figure}

\begin{figure}[tbp]
\centering 
\includegraphics[width=.57\textwidth]{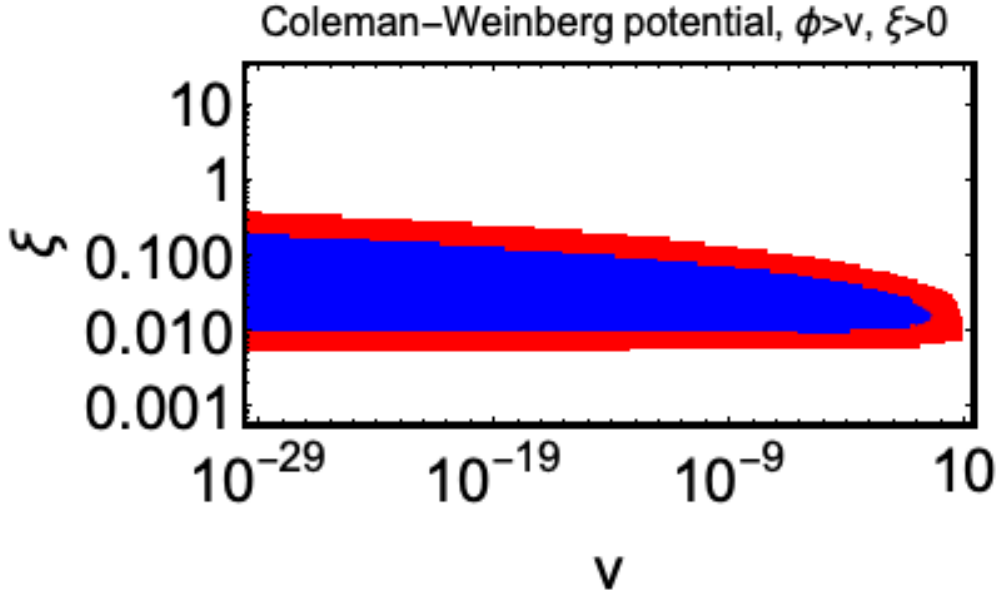}

\

\includegraphics[width=.496\textwidth]{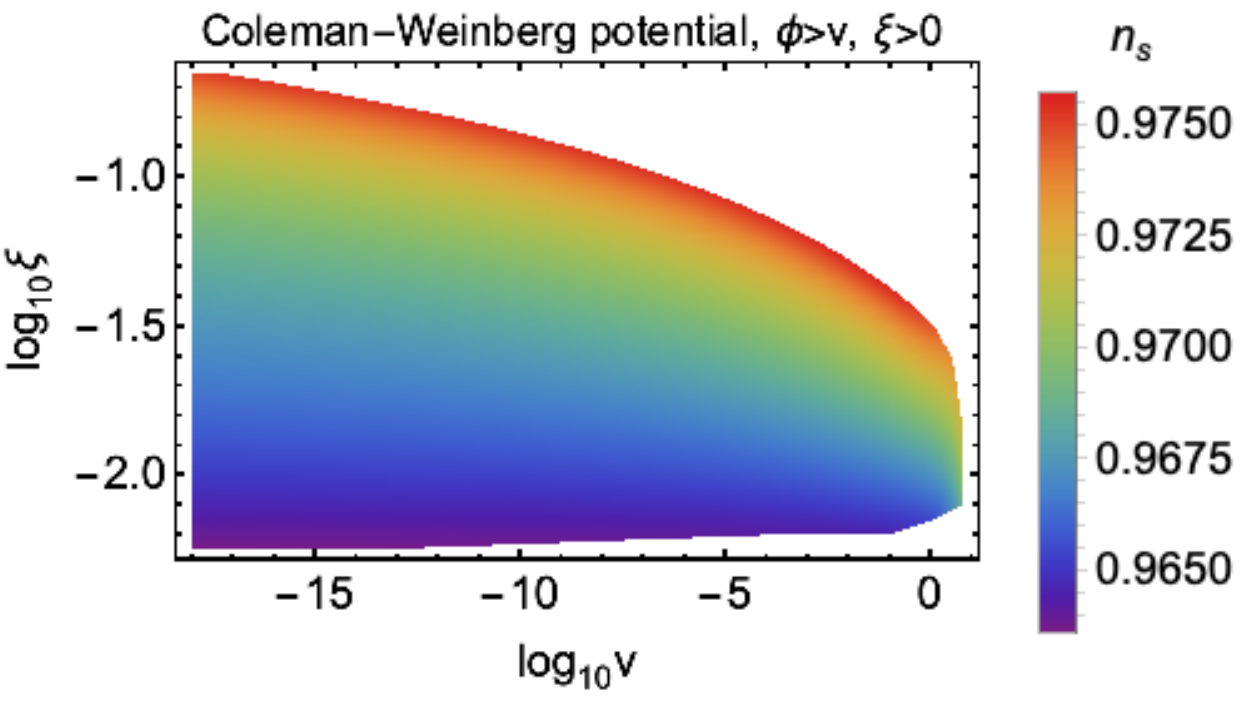}
\includegraphics[width=.496\textwidth]{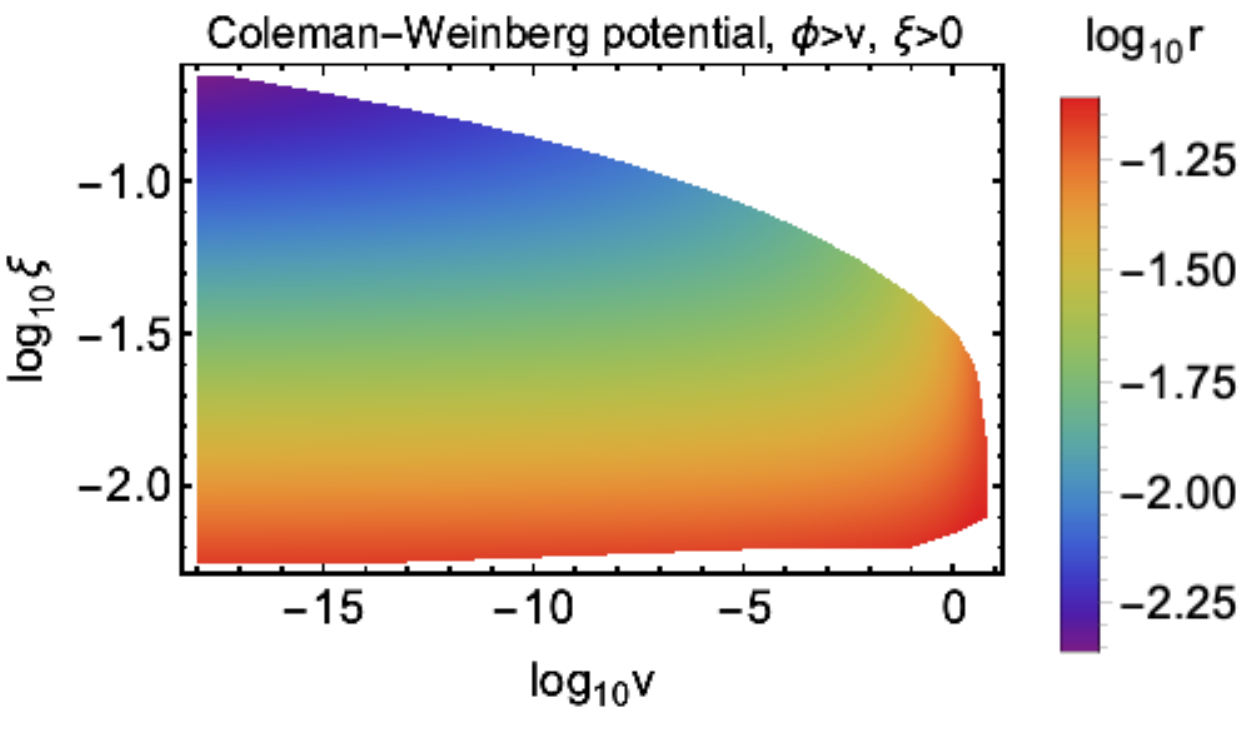}
\caption{\label{cwabovepos} The results of Coleman-Weinberg potential in Palatini formulation for the cases of $\phi>v$ and $\xi>0$. In the top figure, the red (blue) display the regions in the $v-\xi$ plane that predict $n_s$ and $r$ values inside the 95\% (68\%) CL Keck Array/BICEP2 and Planck  contours \cite{Ade:2018gkx}. Bottom figures correspond to the
		$n_s$ and $r$ values in these regions.}
\end{figure}
\begin{figure}[tbp]
\centering 
\includegraphics[width=.58\textwidth]{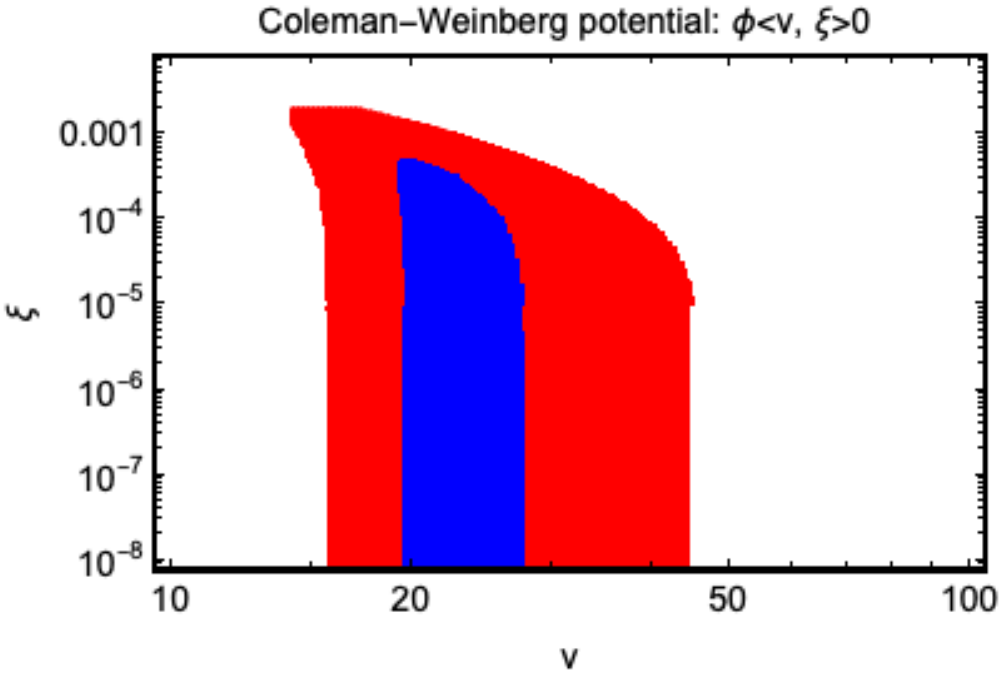}

\

\includegraphics[width=.496\textwidth]{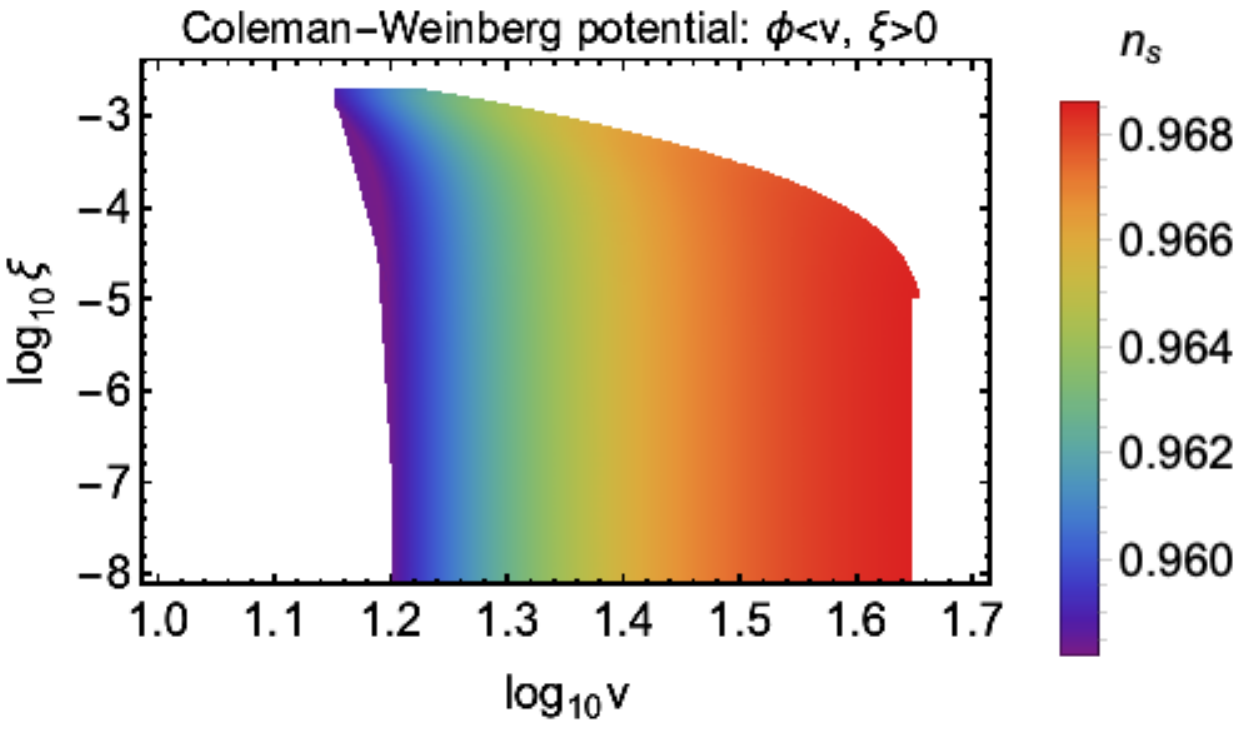}
\includegraphics[width=.496\textwidth]{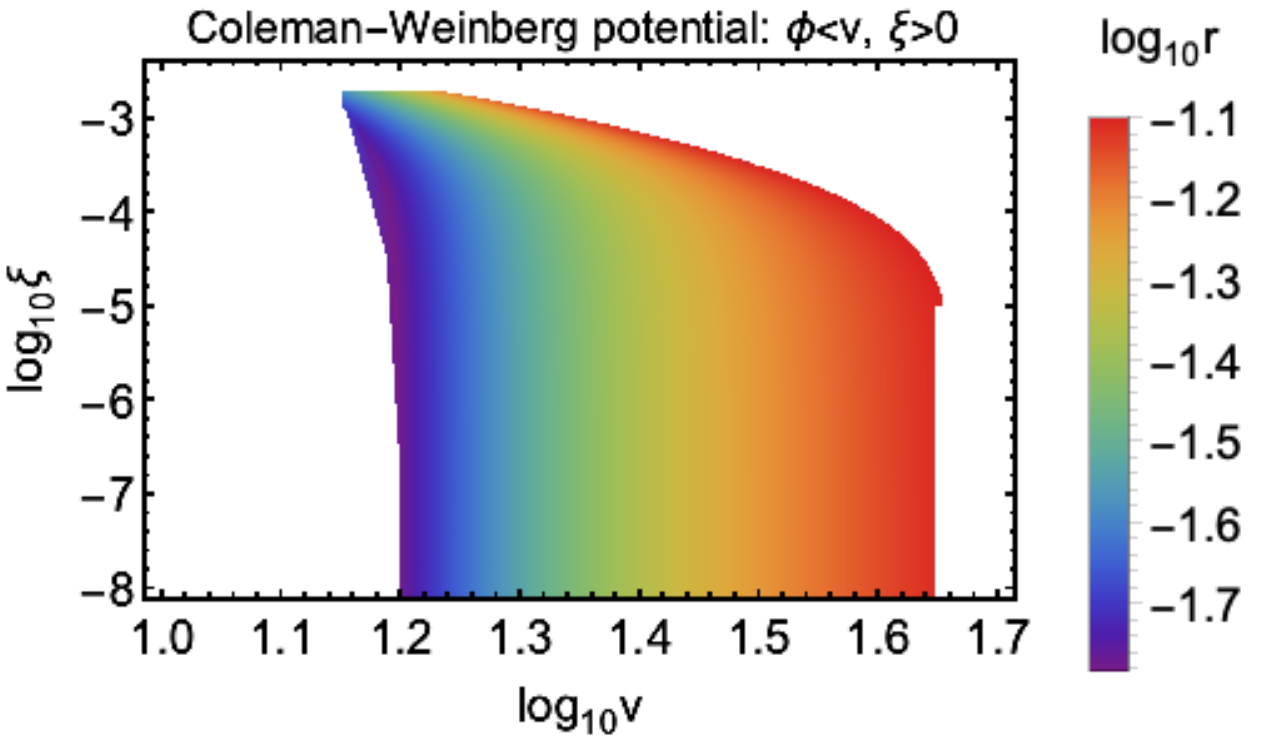}
\caption{\label{cwbelowpos} The results of Coleman-Weinberg potential in Palatini formulation for the cases of $\phi<v$ and $\xi>0$. In the top figure, the red (blue) display the regions in the $v-\xi$ plane that predict $n_s$ and $r$ values inside the 95\% (68\%) CL Keck Array/BICEP2 and Planck  contours \cite{Ade:2018gkx}. Bottom figures correspond to the
		$n_s$ and $r$ values in these regions.}
\end{figure}
\begin{figure}[tbp]
\centering 
\includegraphics[width=.62\textwidth]{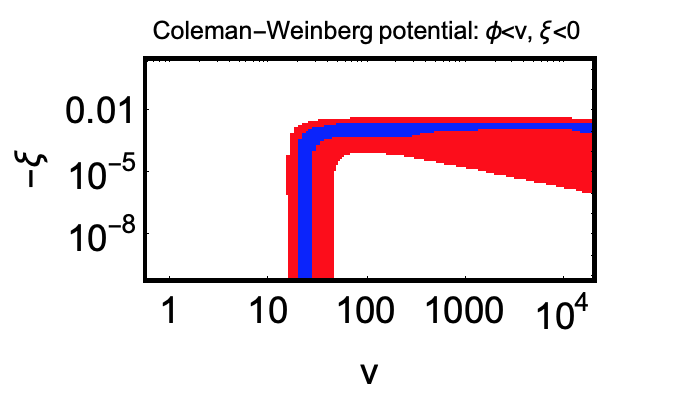}

\

\includegraphics[width=.496\textwidth]{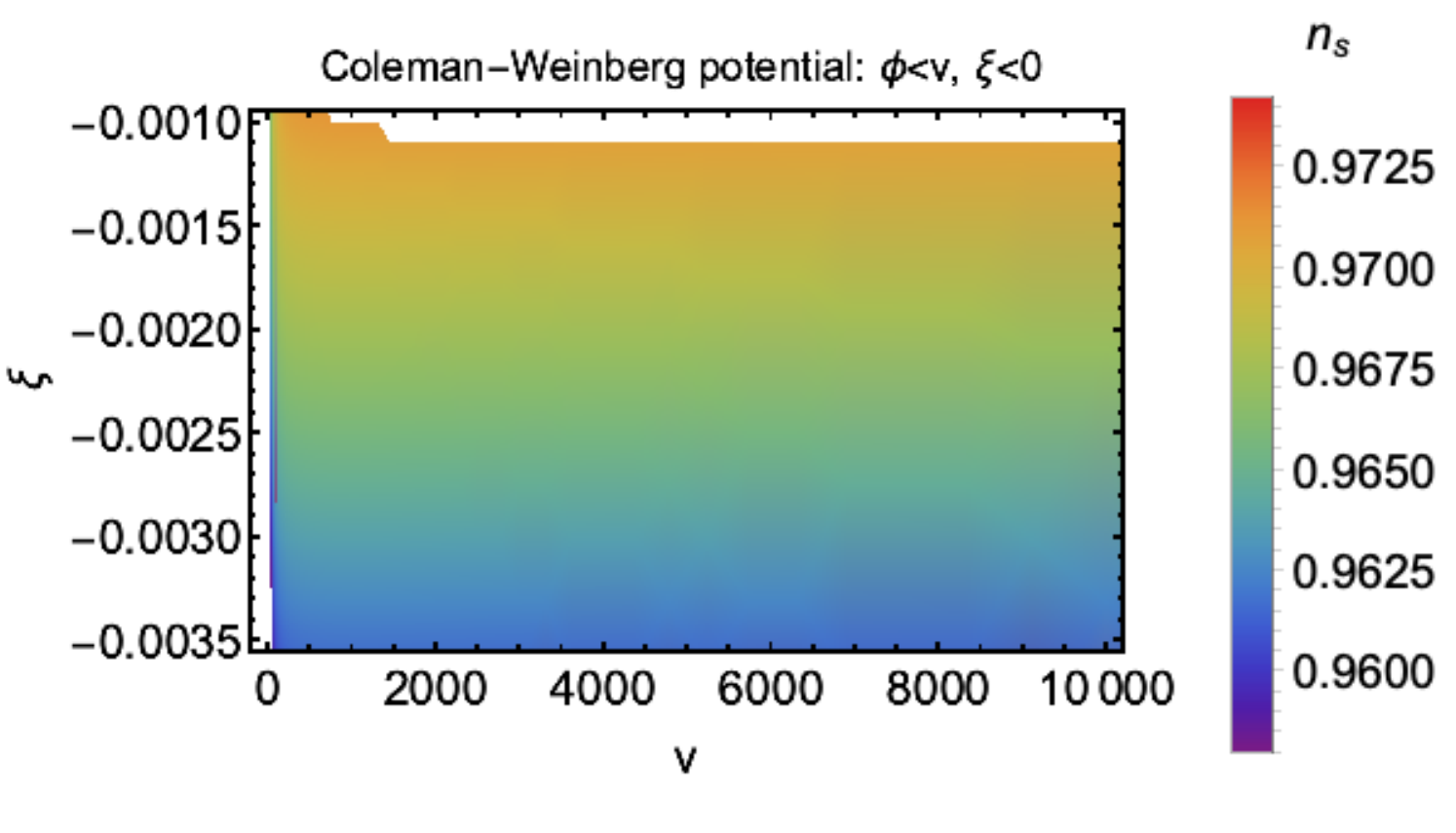}
\includegraphics[width=.496\textwidth]{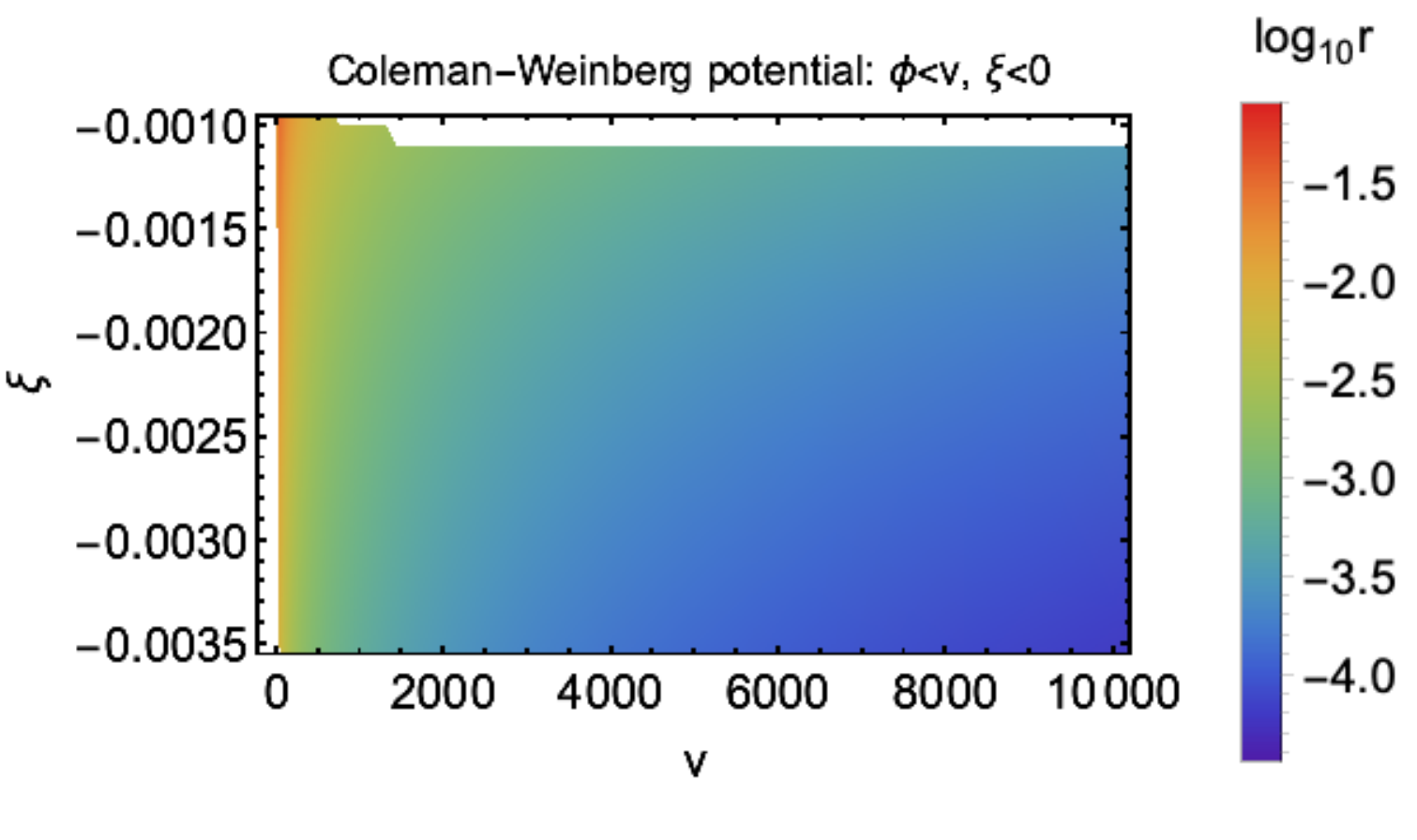}
\caption{\label{cwbelowneg} The results of Coleman-Weinberg potential in Palatini formulation for the cases of $\phi<v$ and $\xi<0$. In the top figure, the red (blue) display the regions in the $v-\xi$ plane that predict $n_s$ and $r$ values inside the 95\% (68\%) CL Keck Array/BICEP2 and Planck  contours \cite{Ade:2018gkx}. Bottom figures correspond to the
		$n_s$ and $r$ values in these regions.}
\end{figure}
\begin{figure}[tbp]
\centering 
\includegraphics[width=.6\textwidth]{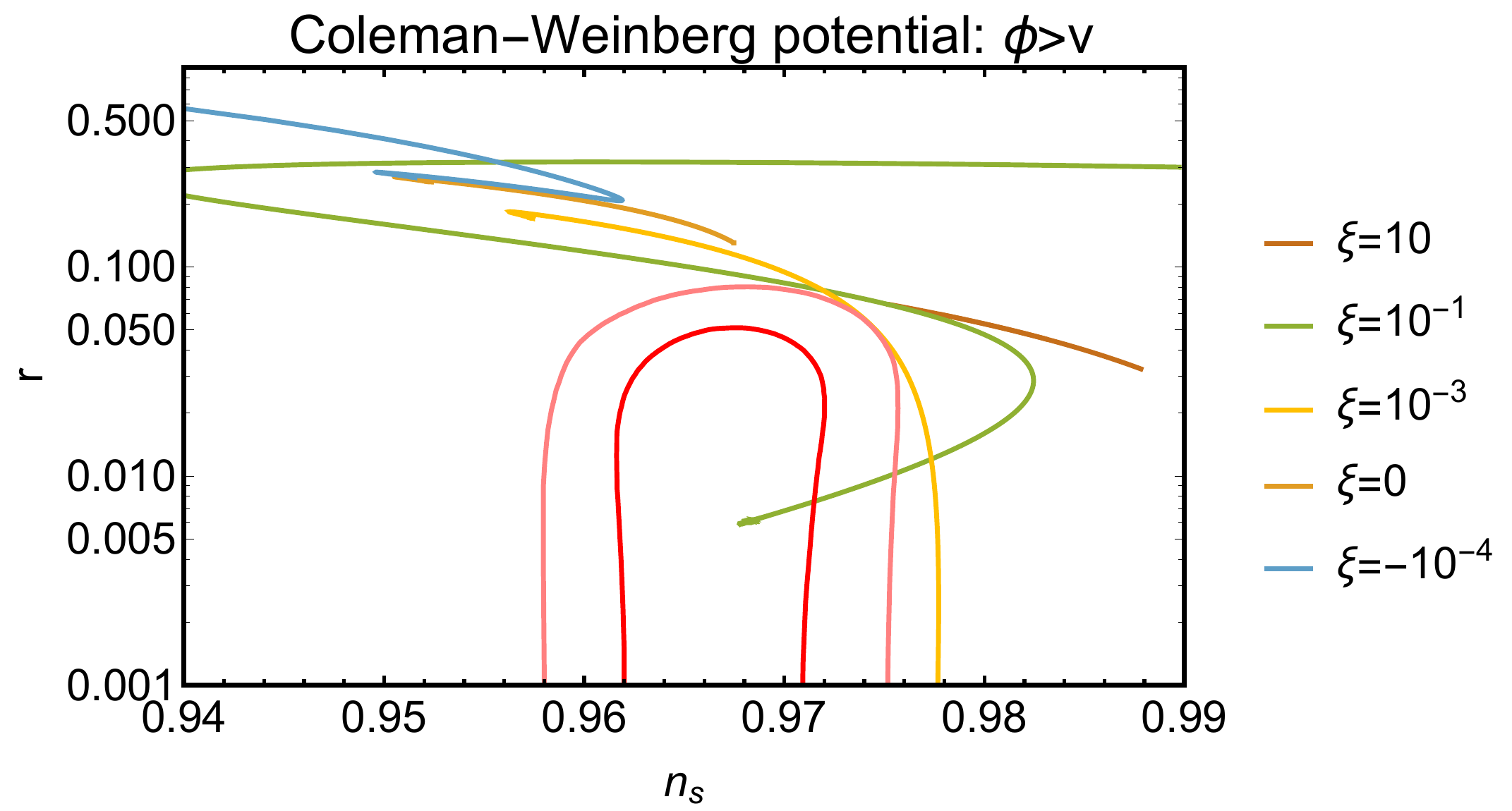}

\

\includegraphics[width=0.96\textwidth]{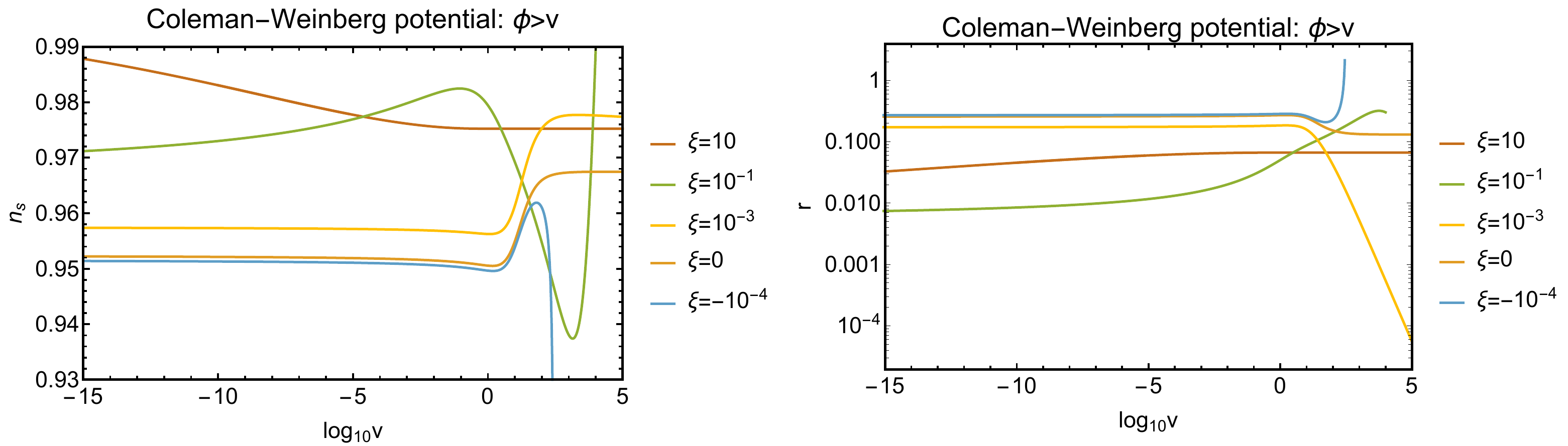}

\caption{\label{cwabovensr} The results of Coleman-Weinberg potential in Palatini formulation for the case of $\phi>v$. Top figure represents the values of $n_s$ and $r$ for different $\xi$ cases as a function of $v$ and the bottom figures indicate the predictions of $n_s-r$ for different $\xi$ values. The pink
		(red) contours indicate the 95\% (68\%) CL Keck Array/BICEP2 and Planck contours  \cite{Ade:2018gkx}.}
\end{figure}
\begin{figure}[tbp]
\centering 
\includegraphics[width=.9\textwidth]{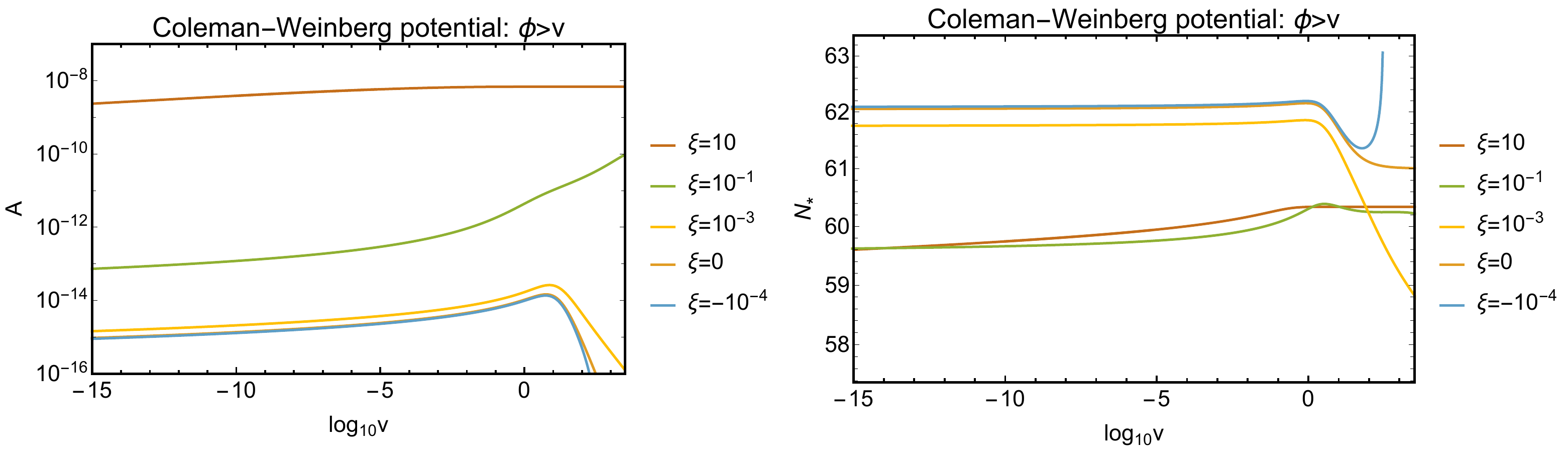}

\

\includegraphics[width=.55\textwidth]{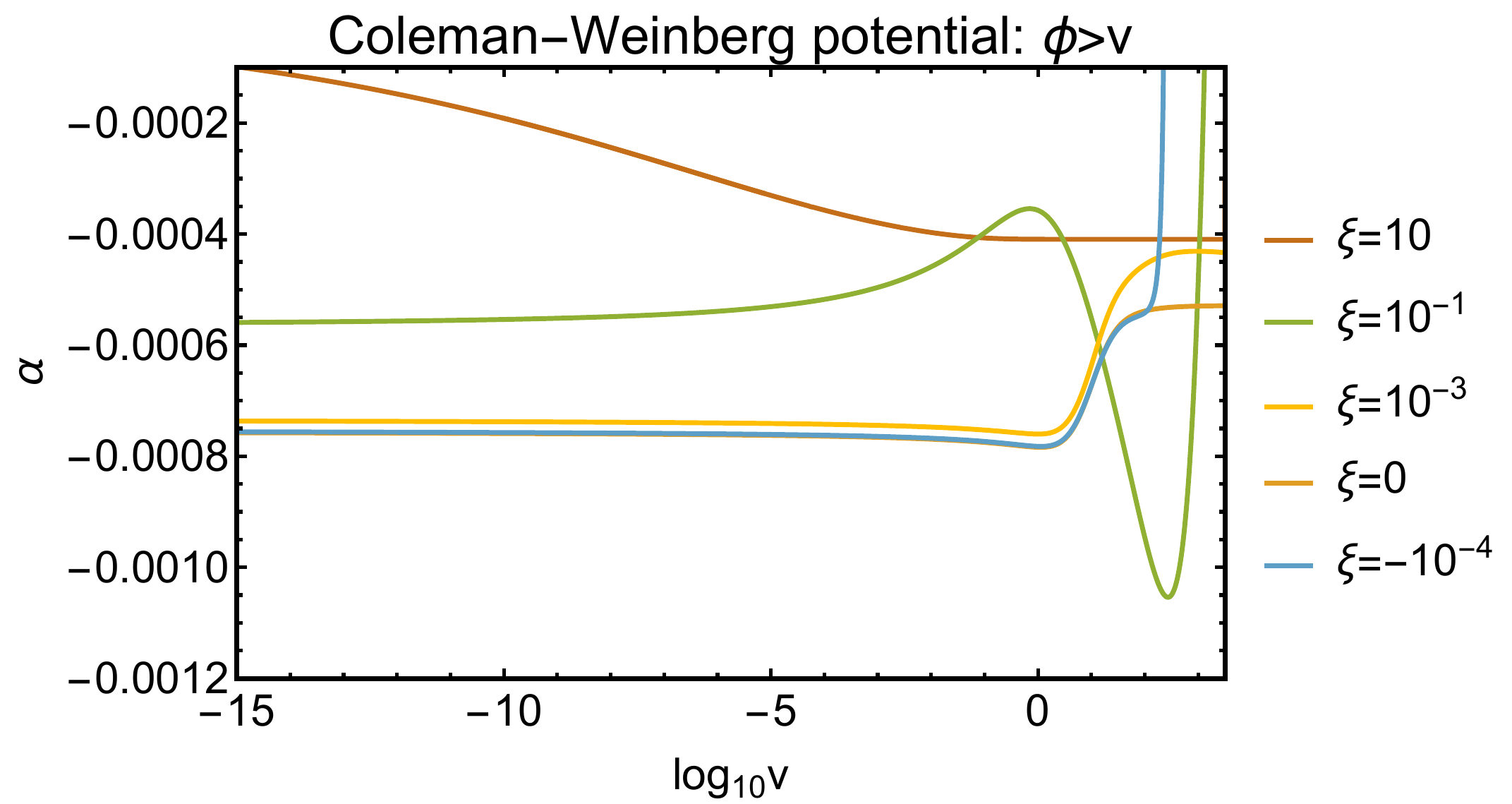}
\caption{\label{cwabovean} The results of Coleman-Weinberg potential in Palatini formulation for the case of $\phi>v$. Top figures show the values of $A$ and $N_*$ as a function of $v$ for different $\xi$ cases and the bottom figure presents the values of $\alpha$ as a function of $v$.}
\end{figure}
\begin{figure}[tbp]
\centering 
\includegraphics[width=.9\textwidth]{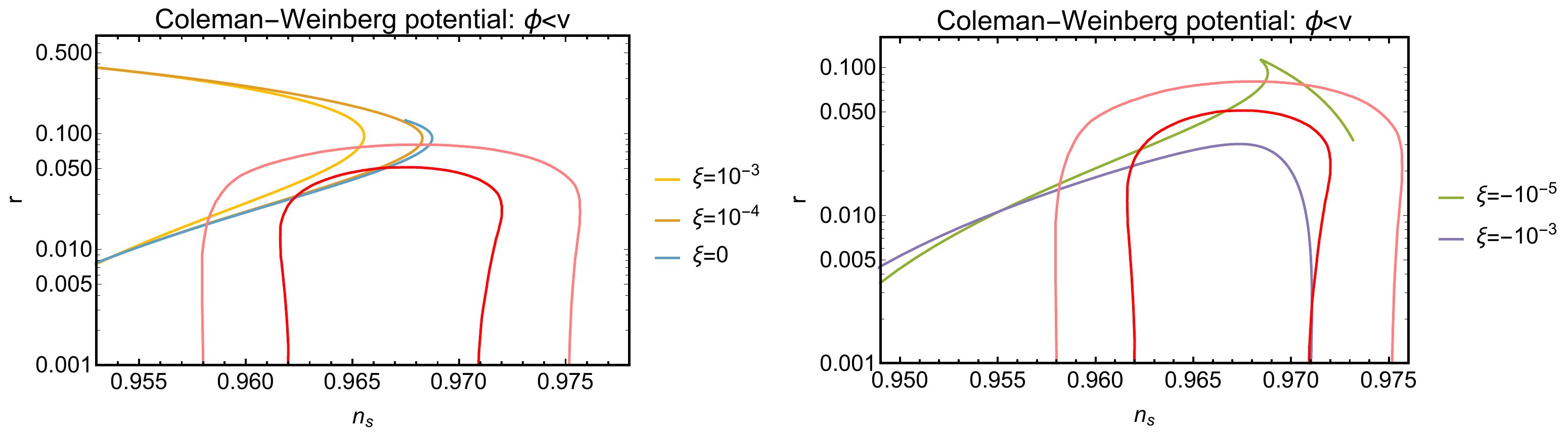}

\

\includegraphics[width=.9\textwidth]{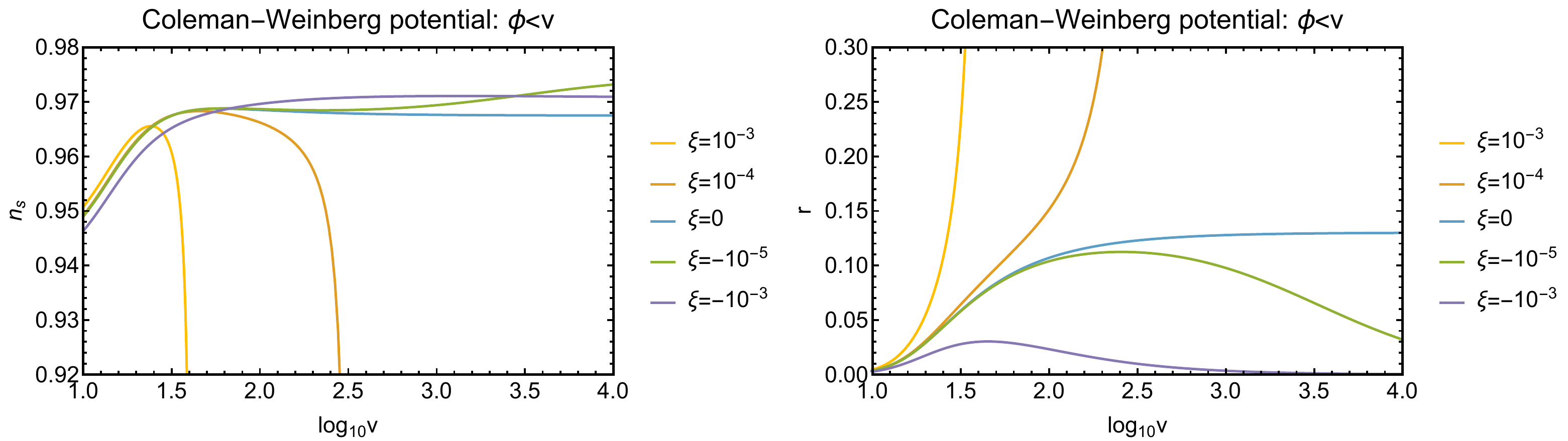}

\caption{\label{cwbelownsr} The results of Coleman-Weinberg potential in Palatini formulation for the case of $\phi<v$. Top figures represent the values of $n_s$ and $r$ for different $\xi$ cases as a function of $v$ and the bottom figures indicate the predictions of $n_s-r$ for different $\xi$ values. The pink
		(red) indicates the 95\% (68\%) CL Keck Array/BICEP2 and Planck contours  \cite{Ade:2018gkx}.}
\end{figure}

\begin{figure}[tbp]
\centering 

\includegraphics[width=.95\textwidth]{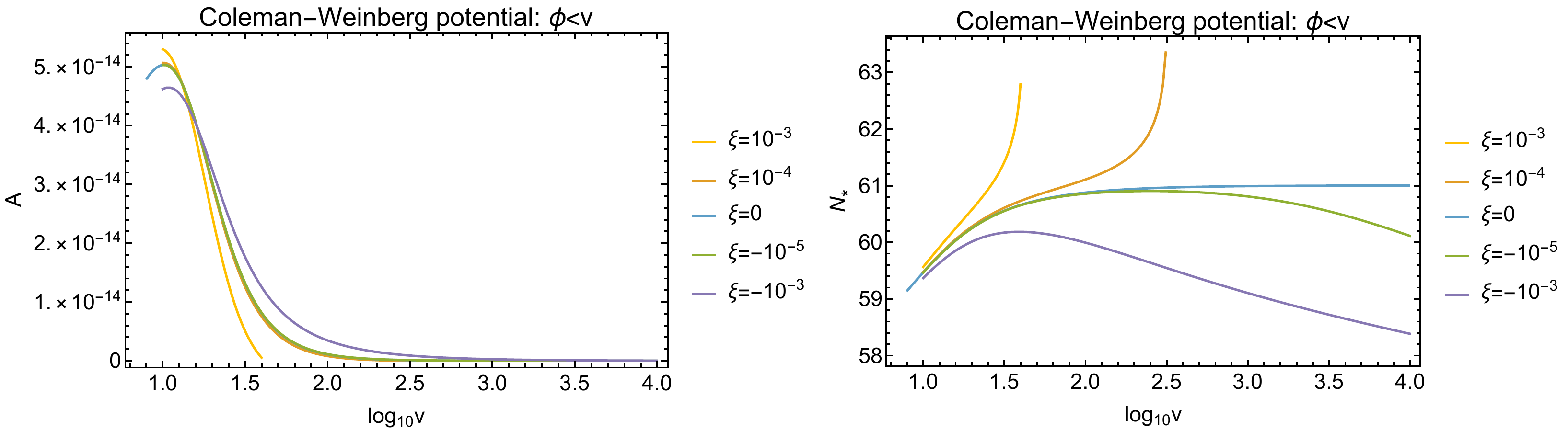}

\

\includegraphics[width=.55\textwidth]{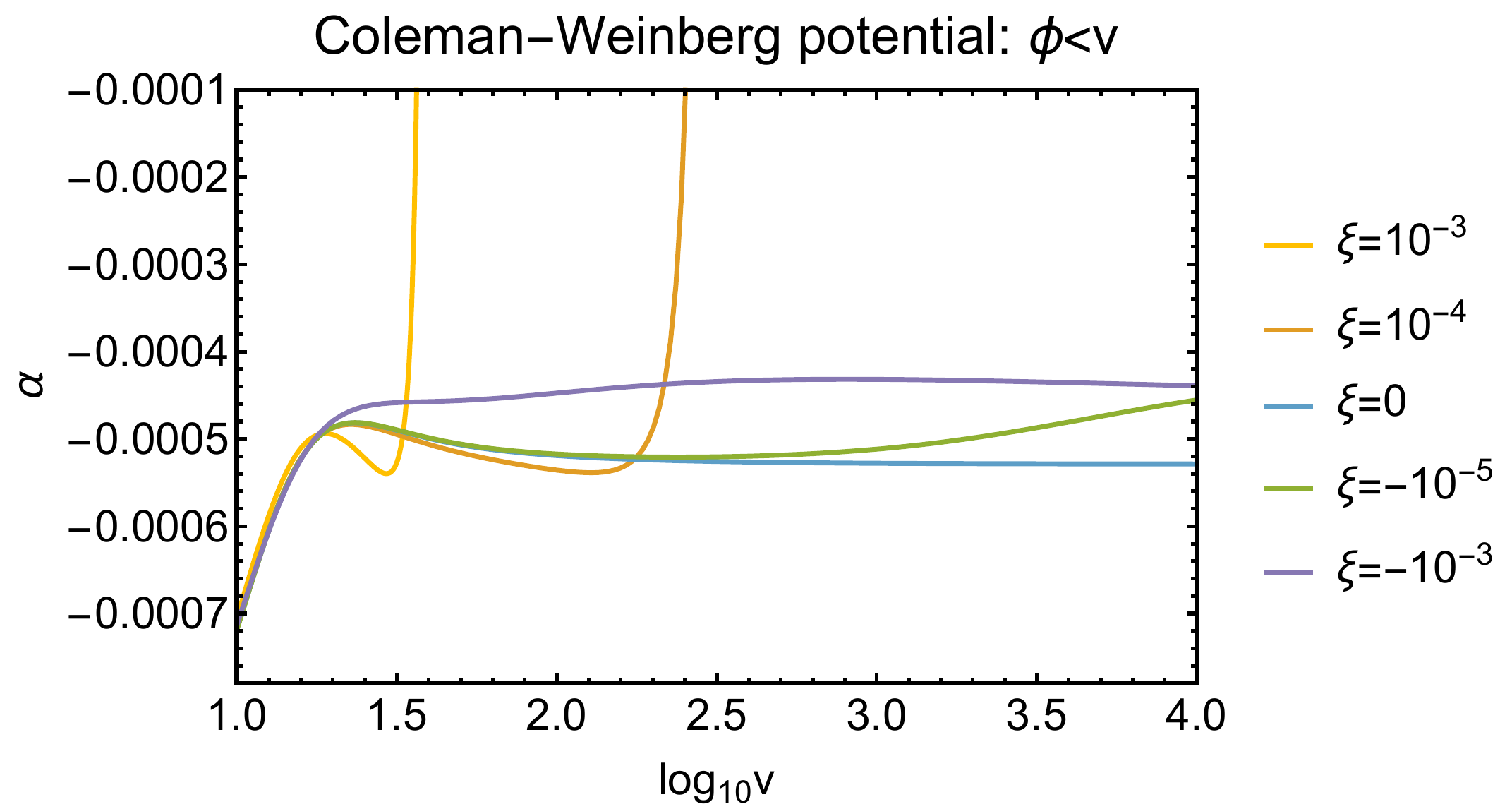}

\caption{\label{cwbelowan} The results of Coleman-Weinberg potential in Palatini formulation for the case of $\phi<v$. Top figures show the values of $A$ and $N_*$ as a function of $v$ for different $\xi$ cases and the bottom figure presents the values of $\alpha$ as a function of $v$.
}
\end{figure}

\

In the induced gravity limit \cite{Zee:1978wi} ($\xi v^2=1$, $F(\phi)=Z(\phi)=\xi \phi^2$), we can obtain the Einstein frame Coleman-Weinberg potential in Palatini formalism by using eq. \eqref{redefine} in the form

\begin{equation}\label{ind}
V(\chi)\simeq \frac{A}{4\xi^2}\left[4\sqrt{\xi}\chi+\exp\left(-4\sqrt{\xi}\chi\right)-1\right].
\end{equation}
During inflation, $\chi>0$. However, for $\phi<v$ inflation, $\chi<0$, so for this case, inflationary parameters are ruled out. For $\phi>v$ inflation, $\chi>0$. In this case, in the $v^2\ll 2N_*$ limits, the potential approximates the linear potential form. This linear behavior of Coleman-Weinberg potential in Palatini formulation agrees with refs. \cite{Racioppi:2017spw, Gialamas:2020snr,Racioppi:2018zoy,Racioppi:2019jsp}. The predictions of Coleman-Weinberg potential in the case of $\xi v^2=1$ are presented in fig. \ref{cwind} for $\xi=1$ and $\xi=10$ values in high-N case. The $n_s-r$ predictions of linear inflation are $n_s\approx 1-3/ 2N_*$ and $r\approx 4/N_*$, thus for $N_*\simeq60$, $n_s\simeq 0.975$ and $r\simeq 0.066$.  As it can be seen from fig. \ref{cwind}, the predictions are ruled out for $\xi=1$ and $\xi=10$ values and when $\xi$ increases, $r$ values decrease and they approximate the linear potential predictions. However, for $\xi \leq 1$ values, the $n_s-r$ can be inside the observational region for the case of $\xi v^2=1$. 

Let's define $\sigma\equiv1/F(\phi)$, so $\sigma=1/1+\xi (\phi^2-v^2)$. We can write $\xi (\phi^2-v^2)\simeq\frac{1}{\sigma}-1\simeq \frac{1}{\sigma}(1-\sigma)$ as well. By using the last definition, this equation can be found
\begin{equation}\label{cwanaly}
\ln\left(\frac{\phi}{v}\right)\simeq \frac{1}{2}\ln \left[1+\frac{1}{\xi v^2 \sigma}(1-\sigma)\right].
\end{equation}
By using eq. \eqref{cwanaly}, the Coleman-Weinberg potential can be written in Einstein frame as follows

\begin{equation}\label{eframe}
V_{E}(\sigma)\simeq \frac{A v^4 \sigma^{2}}{4}\left(\left[1+\frac{1}{\xi v^2 \sigma}\left(1-\sigma\right)\right]^{2}\left(2\ln \left[1+\frac{1}{\xi v^2 \sigma}\left(1-\sigma\right)\right]-1\right)+1\right).
\end{equation}
The Einstein frame Coleman-Weinberg potential, which is defined in eq. \eqref{eframe}, can be simplified by using different approximations and according to these, inflationary predictions can be found for different limit cases. Let's consider that during inflation $\sigma\ll1$ as well as $\big|\frac{1}{\xi v^2 \sigma}\big|\ll 1$. In these limits, we can write the Einstein frame Coleman-Weinberg potential by using $\ln(1+u)\simeq u-\frac{u^2}{2}$, where $u=\frac{(1-\sigma)}{\xi v^2 \sigma}$, as

\begin{equation}\label{eframe2}
V_{E}(\sigma)\simeq \frac{A}{2\xi^2}(1-2\sigma).
\end{equation}
This form of Einstein frame potential approximates the non-minimally coupled Higgs potential. In addition to this, if we consider $u\gg1$ or equivalently $x\equiv\frac{1}{\xi v^2 \sigma}\gg1$ limits, we have solutions just for $\phi>v$ inflation because $\ln\big(1+u\big)$ requires $u>-1$. Furthermore, for these limits, by using $(1+u)\simeq u^2\big(1+\frac{2}{u}\big)$ and $\ln\big(1+u\big)\simeq \ln u+\frac{1}{u}$, Einstein frame Coleman-Weinberg potential can be simplified and the inflationary predictions can be found.

In the literature, \cite{Racioppi:2017spw} discussed the inflationary predictions of Coleman-Weinberg potential in Palatini approach for the case of $v^2=M^2_P/\xi$ and they presented for small values of $\xi $, the predictions approximate the results of quadratic inflation. On the other hand, they also showed while increasing $\xi$, $r$ declines and the predictions coincide with the linear limit for $\xi \gtrsim 0.1$. In this work, we investigate the inflationary predictions of the non-minimally coupled Coleman-Weinberg potential in Palatini formalism for both $\phi>v$ and $\phi<v$ inflation with details. The predictions of this potential are examined in Metric formulation in ref. \cite{Bostan:2018evz}. We present the results of this potential by using high-N case for the regions in the wide $v-\xi$ plane for which the values of $n_s$ and $r$ are inside the 95\% (68\%) CL contour with the data given by the Keck Array/BICEP2 and Planck collaborations \cite{Ade:2018gkx}. We also present the predictions of $n_s$, $r$, $\alpha$, $A$ and $N_*$ for selected $\xi$ values as function of $v$. From figure \ref{cwabovepos}, we see that for $\xi>0$ and $\phi>v$ inflation, predictions are in the observational region for the values of $v\lesssim 1$ and $10^{-2.5}\lesssim \xi \lesssim 10^{-0.5}$.  The values of $v\simeq10^{-15}$ and $\xi\simeq10^{-0.5}$, $n_s\simeq 0.975$ and $r\simeq 0.006$ but for $v\simeq10^{-15}$ and $\xi\simeq 10^{-2.5}$, $n_s\simeq 0.965$ and $r\simeq 0.06$. Furthermore, in figure \ref{cwbelowpos}, which shows the results of $\xi>0$ and $\phi<v$ inflation, we can see that the $n_s-r$ values are in the 68\% CL contour for the range of $20\lesssim v \lesssim 40 $ and $10^{-5} \lesssim \xi \lesssim 5\times 10^{-4}$. For $\xi\simeq 10^{-2.8}$ and $17\lesssim v \lesssim 21$, the predictions are in 95\% CL contour and $n_s\simeq0.96$, $r\simeq0.025$. For the cases of $\xi<0$ and $\phi<v$, predictions can be inside the observational region, as it can be seen from fig. \ref{cwbelowneg}.

\

Furthermore, in this section, we present the values of $n_s, r, A, N_*, \alpha$ for the non-minimally coupled Coleman-Weinberg potential in Palatini formalism as a function of $v$ for different $\xi$ values which we selected. As it can be seen from fig. \ref{cwabovensr}, for $\phi>v$ inflation, just for $\xi=10^{-1}$, the predictions can be inside the observational region for $v\ll 1$. On the other hand, for another set of selected $\xi$ values, the predictions are ruled out. Fig. \ref{cwabovean} shows the results of $A, N_*$ and $\alpha$ for $\phi>v$ inflation for different $\xi$ values. We can see that the prediction of $\alpha$ is too small to be measured in the forthcoming experiments. In addition to this, we also examined the predictions for $\phi<v$ inflation and we presented the results in fig. \ref{cwbelownsr}. According to this figure, the predictions can be in the observational region for $\xi=10^{-3},\xi=10^{-4}$, $\xi=0$ and $\xi=-10^{-5}$, as well as $\xi=-10^{-3}$ values. Fig. \ref{cwbelowan} presents the results of $A, N_*$ and $\alpha$ for $\phi<v$ inflation for selected $\xi$ values. Similar to the $\phi>v$ inflation, the value of $\alpha$ is very tiny to be measured in the upcoming experiments.
\clearpage
\section{Conclusion}\label{conc}
In this work, we analyzed the inflationary parameters of two symmetry breaking type potentials: Higgs potential and Coleman-Weinberg potential in Palatini formulation. We considered the inflaton has a non-zero $v$ after the period of inflation, and the non-minimal coupling function is $F(\phi)=1+\xi(\phi^2-v^2)$. For each potential, the regions in the $v-\xi$ plane, we showed the predictions of the inflationary parameters $n_s$ and $r$ are in agreement with the recent measurements. We also presented the inflationary predictions of Coleman-Weinberg potential for preferred $\xi$ values as a function of $v$. 

We obtained $n_s\simeq 0.974$ and $r\simeq 10^{-15}$ for the Palatini Higgs inflation, for the values of $\xi, v\gg1$ and the predictions are in the 95\% CL contour in this limit. We can say that for the large-field limit, $r$ is very suppressed in Palatini Higgs inflation for the case of $\phi>v$. However, for $\phi<v$ and $\xi>0$ inflation, the predictions are in the observational region for just the range of $10^{-5}\lesssim \xi \lesssim 10^{-3}$ and $17\lesssim v \lesssim 32$ and their values are $n_s\simeq 0.965$ and $0.032\lesssim r \lesssim 0.079$. We also analyzed the cases of $\phi<v$ and $\xi<0$ inflation, and  the predictions of $n_s$ and $r$ can be inside the 68\% and 95\% CL. For the values of $-100\lesssim \xi \lesssim -10^{-3}$ and $10\lesssim v \lesssim 30$, the predictions are in 68\% CL but for the same $\xi$ ranges and $30\lesssim v \lesssim 10^{4}$, $n_s$ and $r$ are in 95\% CL. 

Furthermore, we analyzed the predictions of Coleman-Weinberg Palatini inflation both analytically and numerically. We observed for the $\xi v^2=1$ limit, the $n_s-r$ predictions are ruled out for $\xi=1$ and $\xi=10$ values and when $\xi$ increases, $r$ decreases and the predictions saturate the linear potential limit. We also showed that we only have a solution for $\phi>v$ inflation in $\xi v^2=1$ limit because of $\chi>0$ during inflation. In addition to this, we presented for Coleman-Weinberg $\phi>v$ and $\xi>0$ inflation, the predictions are in agreement with the recent data for the values of $v\lesssim1$ and $10^{-2.5}\lesssim \xi \lesssim 10^{-0.5}$. We found $n_s\simeq0.975$ and $r\simeq 0.006$ for the limits of $\xi\simeq10^{-0.5}$ and $v\ll1$. Furthermore, we also presented the results of $\xi>0$ and $\phi<v$ inflation, $n_s-r$ predictions are in the 68\% CL contour for $20\lesssim v \lesssim 40 $ and $10^{-5} \lesssim \xi \lesssim 5\times 10^{-4}$ region. For $\xi\simeq 10^{-2.8}$ and $17\lesssim v \lesssim 21$, the predictions are in 95\% CL contour and $n_s\simeq0.96$, $r\simeq0.025$. For $\xi<0$ and $\phi<v$ inflation, we obtained the predictions can be inside the observational region.

Finally, we presented the values of $n_s, r, A, N_*,\alpha$ for the Coleman-Weinberg Palatini inflation as a function of $v$ for preferred $\xi$ values. It can be specially emphasized that we observed the predictions of $\alpha$ is too small for the considered potential to be observed in the upcoming measurements.

\newpage

\acknowledgments
The author would like to thank Antonio Racioppi, Alexandros Karam and O. K. Koseyan for their valuable comments.



\end{document}